\documentclass[superscriptaddress,reprint,amssymb,amsmath,aps,prd]{revtex4-1}
\usepackage{graphicx}
\usepackage{bm}
\usepackage{bbm}
\usepackage{amsthm}
\usepackage{braket}
\usepackage{amsfonts}
\usepackage{graphicx}
\usepackage{hyperref}
\usepackage{amsthm}

\begin{document}
\title{History state formalism for scalar particles} 
\author{N. L.\  Diaz}
\affiliation{Departamento de F\'isica-IFLP/CONICET,
	Universidad Nacional de La Plata, C.C. 67, La Plata (1900), Argentina}

\author{J. M. Matera}
\affiliation{Departamento de F\'isica-IFLP/CONICET,
	Universidad Nacional de La Plata, C.C. 67, La Plata (1900), Argentina}

\author{R. Rossignoli}
\affiliation{Departamento de F\'isica-IFLP/CONICET,
	Universidad Nacional de La Plata, C.C. 67, La Plata (1900), Argentina}
\affiliation{Comisi\'on de Investigaciones Cient\'{\i}ficas (CIC), La Plata (1900), Argentina}

\begin{abstract}
    We present a covariant quantum formalism for scalar particles based on an enlarged Hilbert space. The particular physical theory can be introduced through a timeless Wheeler DeWitt-like equation, whose projection onto four-dimensional  coordinates leads to the Klein Gordon equation. The standard quantum mechanical product in the enlarged space, which is invariant and positive definite,  implies the usual Klein Gordon product when applied to its eigenstates. Moreover, the standard three-dimensional invariant measure emerges naturally from the flat measure in four dimensions  
    when mass eigenstates are considered, allowing a rigorous identification between definite mass history states and the standard Wigner representation. Connections with the free propagator of scalar field theory and localized states are subsequently derived. The formalism also allows  the superposition of different theories and remains   valid in the presence of a fixed external field, revealing special orthogonality relations. Other details such as extended identities for the current density, 
    the quantization of parameterized theories and the non relativistic limit, with its connection to the Page and Wooters formalism, are discussed.
    A related consistent second quantization formulation is also introduced.  
\end{abstract}

\maketitle

\section{Introduction}

The introduction of the concept of time in a quantum mechanical framework \cite{PaW.83,g.09,m.00} has recently attracted renewed attention \cite{QT.15,b.16,m.17,e.17,d.17, b.18,m.18,m.19,di.19,sm.19}.
One persistent motivation is its connection with fundamental open problems, mainly related to the 
quantization of gravity \cite{d.58,ma.95,ma2.95,ma.97,ma.00,qg,p.99,ku.11,b.11,r.11,b.12,a.12,chat.19}, whose classical 
description is a general covariant theory \cite{e.16}. 
While the Page and Wootters (PaW) formalism 
\cite{PaW.83} has been able to provide 
a successful quantum treatment of time \cite{g.09,QT.15}, it was mainly exploited to obtain non relativistic equations, namely, the Schr\"odinger equation \cite{QT.15} and its discretized version \cite{b.18}.
However, the rigorous definition of an hermitian time operator, enabled by this formalism through an enlarged Hilbert space, 
has opened the possibility to explore the construction of explicitly covariant representations. This idea  was recently employed to embed the Dirac equation \cite{di.28} within a covariant Hilbert space formalism  \cite{di.19}.

In this work we exploit these concepts further and develop the case of scalar particles, gaining new insight on the subject. 
One of the main results is the definition of a consistent Hilbert space for the Klein Gordon equation  \cite{K.26,G.26}, in both the free case and in the presence of an external field, where the inner product is the canonical product in four  dimensions. 
Remarkably, this construction, and the subsequent proper normalization of fixed mass states, which are eigenstates of a Wheeler DeWitt-like equation \cite{wdw.67}, ensure the usual three-dimensional ($3d$) norm. 
Moreover, in the free case, the subspace of definite mass maps onto the standard Wigner representation \cite{w.39}, directly implying the standard $3d$ invariant measure. 
 While corresponding results for the free case were previously obtained in the context of quantum gravity  
\cite{ma.97,ma.00,ma.95,ma2.95}, the four-dimensional ($4d$) space was there considered as an auxiliary (kinematic) Hilbert space (from which the important result of an induced $3d$ product for ``physical'' states was inferred). Here we promote it to the status of a real physical space. This allows one to upgrade time from a  parameter to an operator, which in turn requires to promote mass, which in both Dirac and Klein Gordon equations is assumed as a fixed parameter, to a quantum observable.  This approach offers substantial conceptual advantages even if just the subspace (eigenspace) of definite mass states is  considered, but in addition it opens the way to new possibilities \cite{di.19}, such as more general quantum states with mass fluctuations 
and an extended Fock space based on four dimensional entities. Moreover, the present treatment of interactions reveals that such general states are already implied when expressing the corresponding solutions in terms of the free states, 
in analogy with the off-shell contributions in perturbative treatments for interacting many particle 
systems.
These results provide a new perspective which could be suitable to deal with the Hilbert space problem of the Wheeler DeWitt framework of quantum gravity \cite{wdw.67,ku.11,a.12,b.11}. 

The basic construction of the explicitly covariant Hilbert space adequate for scalar particles is presented in Sec.\ \ref{sec:scalarpa},  
where event states $|x\rangle$ are defined as eigenstates of the hermitian operators $X^\mu$, with $X^0$ introduced in accordance with the PaW formalism. 
It is then shown that the $3d$  Klein Gordon product emerges from the $4d$ 
orthogonality of mass eigenstates.
This leads to Sec.\ \ref{sec:wigner} where the relation with the standard single particle representation of the Poincar\'e group \cite{w.39} is established, together with the one to one correspondence between the $4d$ fixed mass history states and those of the usual scalar Wigner representation. 
Since the history states are more general, 
 this correspondence only holds in a particular mass subspace, excluding thus the states $|x\rangle$.  Yet, it is shown in Sec.\ \ref{sec:prop} that the space-time localized states can be projected onto the ``physical subspace'' providing geometrical physical information. This result is employed to obtain the free propagation amplitude of scalar field theory \cite{p.18} within the present formalism. 
The proper action of covariant operators on physical subspaces is further clarified in Sec.\ \ref{sec:normtime} by decomposing the Hilbert space according to its different mass and energy sectors. In particular, the unboundedness of $P^0$, the generator of the time translations, is discussed. The normalization in time is considered in the same section, where it is explicitly shown that a general normalizable state in the covariant Hilbert space is a superposition of the previous mass ``improper'' eigenstates.

In Sec.\ \ref{sec:ext} the universe equation is generalized to include 
interactions with an external field. The Klein Gordon equation with a potential is obtained by projecting onto $|x\rangle$ the associated eigenvalue equation. It is then proved that the correct connection between the canonical extended product and the Klein Gordon product holds for any mass and time independent external field (for a given gauge choice and reference frame). It is also remarked how the consideration of states with no definite mass is already implicit when dealing with interactions. 

Some of the new insights which follow from the relativistic regime are transferred to the non relativistic case in Sec.\ \ref{sec:nonrel}. In particular, a proposal for the normalization of states with infinite histories is derived in a self-contained non relativistic discussion. The case of a linearly mass dependent potential is also briefly discussed.

The consistent construction of the single particle representation also allows a consistent definition of a Fock space where the building block is the particle as a four-dimensional entity. In Sec.\ \ref{sec:fock} this ``second quantization of histories'' is explored. The identification of Sec.\ \ref{sec:wigner} is extended to the standard Fock space of scalar field theory through the definition of a proper subspace and the generalization of the universe operator to a one-body operator. Finally, conclusions and perspectives are discussed in Sec.\ \ref{sec:disc}.

\section{Scalar Particle}\label{sec:scalarp}
\subsection{Quantum Formalism}\label{sec:scalarpa}
	A general history state 
	for a scalar particle 
	can be written as 
			\begin{equation}
	|\Psi\rangle=\int d^{4}p \;\Psi(p)|p \rangle
	\end{equation}
	where $|p\rangle\in {\cal H}$ are the improper eigenstates of the four operators $P_{\mu}$. 
	Here ${\cal H}=\{S(\mathbb{R}^4),L^2(\mathbb{R}^4),S^*(\mathbb{R}^4)\}$ is the rigged Hilbert space constructed from $L^2(\mathbb{R}^4)$, and $S(\mathbb{R}^4)$ is the Schwartz space. 
	Boost operators are defined by 
	\begin{equation}
	U(\Lambda)|p\rangle=|\Lambda p\rangle, \label{ust}\end{equation}
	with $ \Lambda^{\mu}_{\;\nu}=e^{w^{\mu}_{\;\nu}}$ and $w_{\mu\nu}=-w_{\nu\mu}$ an antisymmetric tensor. 
	The transformed state 
	becomes 
	\begin{equation}
	U(\Lambda) |\Psi \rangle= \int d^{4}p\, \Psi'(p)|p\rangle\,,
	\end{equation}
	with \begin{equation}\label{eq:transformation}  \Psi'(p)= \langle p|U(\Lambda)|\Psi \rangle=\Psi(\Lambda^{-1}p)\,.
	\end{equation}
	We may also introduce the states 
	$|x\rangle=\frac{1}{(2\pi)^2}\int  d^4p\, e^{ipx}|p\rangle$ with $px=p_\mu x^\mu=p^0x^0-\sum_{i=1}^3 p^i x^i$, which transform as $U(\Lambda)|x\rangle=|\Lambda x\rangle$. If $|x\rangle$ are eigenstates of operators $X^{\mu}$, the latter satisfy the  commutation relations $\left[X^{\mu},P_{\nu}\right]=i \delta^{\mu}_{\;\nu}$. Clearly the operators $P_\mu$, $L_{\mu\nu}:=X_\mu P_\nu-X_\nu P_\mu$ provide a representation of the Lie algebra of the Poincar\'e group, where it is worth noting that $P_0$ is not the Hamiltonian (see Sec.\ \ref{sec:normtime}) and that the representation acts on $\mathcal{H}$ and not on a classical field. The representation is manifestly unitary since 
	\begin{equation}
	    \langle \tilde{\Psi}|U(\Lambda)^\dag U(\Lambda)|\Psi\rangle=  \int d^4p\, \tilde{\Psi}'^\ast(p)\Psi'(p)=\langle \tilde{\Psi}|\Psi\rangle\,.
	\end{equation}
	
		Next we consider the operator
	\begin{equation}
	{\cal J}=P^\mu P_\mu\label{JD}\,.
	\end{equation} 
	The equation  
	\begin{equation}\label{eqgd}
	{\cal J}|\Psi\rangle=m^2|\Psi\rangle\,, 
	\end{equation} 
	has the general solution 
	\begin{eqnarray}\label{eq:solutions}
	|\Psi_{m^2}\rangle&=&
	\int d^{4}p \; \delta(p^\mu p_\mu-m^2)H^+(p^0) \alpha(\textbf{p})|p\rangle 
	\nonumber\\&&\oplus 
	\int d^{4}p \; \delta(p^\mu p_\mu-m^2)H^-(p^0)\beta(\textbf{p})|p\rangle
	\end{eqnarray}
	where $H^\pm$ denotes the Heaviside function such that $\pm$ corresponds to positive 
	or negative $p^0$ and $m^2$ is a real eigenvalue of the hermitian operator $\mathcal{J}$.

Defining $\Psi(x)=\langle x|\Psi\rangle$, Eq.\ (\ref{eqgd}) becomes the usual Klein Gordon equation \cite{K.26,G.26}, 
    \begin{equation}\label{eq:dirac}
	\langle x|(P^\mu P_\mu-m^2)| \Psi\rangle=0\; \Rightarrow\; 
	(\partial^\mu\partial_\mu +m^2)\Psi(x)=0\,
	\end{equation} 
	whose invariance is apparent since $\Psi'(x)=\langle x|U(\Lambda)|\Psi\rangle=\Psi(\Lambda^{-1}x)$. 
	Since 
	\begin{equation}
	    \delta(p^{\mu}p_{\mu}-m^2)H^{+}(p^0)=\frac{\delta(p^0-E_{\textbf{p}m})}{
	    2E_{\textbf{p}m}}\,,
	\end{equation} 
	with $E_{\textbf{p}m}=\sqrt{\textbf{p}^2+m^2}$, 
	an arbitrary solution with positive $p^{0}$ can be written explicitly as 
\begin{eqnarray}
	|\Psi_{m^2}\rangle&=&\frac{1}{\sqrt{2\pi}}\int d^{4}x\; \psi(x)|x\rangle\,,\label{eq:psix0} \\ \psi(x)&=& \frac{1}{(2\pi)^{3/2}}\int \frac{d^{3}p}
	{2 E_{\textbf{p} m}}\alpha(\textbf{p})
	e^{-ipx|_{p^0=E_{\textbf{p} m}}}\,,
	\label{eq:psix}\end{eqnarray}
	 where $\psi(x)=\sqrt{2\pi} \Psi(x)$. Under a Lorentz transformation, $\alpha(\textbf{p})\to \alpha(\Lambda^{-1}\textbf{p})$ (Eq.\ (\ref{eq:solutions})), implying 
				 $\frac{d^3p}{2E_{\textbf{p}m}}$ invariant, in agreement with the well known result. 
	The product of two solutions 
	corresponding to different eigenvalues $m^2$ and $\tilde m^2$ yields  	
	\begin{eqnarray}
	\langle \Psi_{\tilde m^2}|\Psi_{m^2}\rangle&=&\int \frac{d^3p}{4E_{\textbf{p}\tilde m}E_{\textbf{p}m}}\delta(E_{\textbf{p} m}-E_{\textbf{p}\tilde m})
	 \tilde{\alpha}^*(\textbf{p})\alpha(\textbf{p})\;\;\; \nonumber\\
	&=&\delta(m^2-\tilde m^2)\int \frac{d^3p}{2E_{\textbf{p}m}}
	\tilde{\alpha}^*(\textbf{p})\,\alpha(\textbf{p}) \label{eq:deltanorm0}
	\end{eqnarray} 
	since $\delta(E_{\textbf{p} m}-E_{\textbf{p}\tilde m})=\delta(m^2-\tilde{m}^2)2E_{\textbf{p}m}$. 
	In the case of two solutions with the same momenta distribution at equal mass, then 
\begin{eqnarray}
	\langle \Psi_{\tilde m^2}|\Psi_{m^2}\rangle&=&\delta(m^2-\tilde m^2)\int \frac{d^3p}{2E_{\textbf{p}m}} |\alpha(\textbf{p})|^2\, \label{eq:deltanorm}
	\end{eqnarray} 
with a similar expression in terms of $\beta(\bf{p})$ for negative $p^0$ (solutions with positive and negative $p^0$ are orthogonal). \\
It is straightforward to see from Eq.\ (\ref{eq:psix}) that  \[\int\frac{d^3p}{2E_{\textbf{p}m}} |\alpha(\textbf{p})|^2=Q(\psi,\psi)\] 
with
\begin{equation}
    Q(\varphi,\psi):=i\int d^3x \; \left(\varphi^\ast(\textbf{x},t)\partial_t\psi(\textbf{x},t)-\psi(\textbf{x},t)\partial_t\varphi^\ast(\textbf{x},t)\right)
\end{equation}
and $\psi(\textbf{x},t)=\psi(x)$. 
Since \begin{equation}
    \langle \Psi_{\tilde m^2}|\Psi_{m^2}\rangle=\delta(m^2-\tilde m^2)\,Q(\psi,\psi)\,,\label{qpsi}
\end{equation} 
the proper normalization of these
solutions in $S^\ast(\mathbb{R}^4)$ then implies, remarkably, the usual Klein Gordon normalization \cite{K.26} $Q(\psi,\psi)=1$, i.e.
\begin{eqnarray}\label{eq:norm}
&&\hspace*{-0.75cm}\langle \Psi_{\tilde m^2}|\Psi_{m^2}\rangle=\delta(m^2-\tilde m^2)\, \Leftrightarrow \nonumber\\ &&\hspace*{-0.75cm}i\int d^3x \; \left(\psi^\ast(\textbf{x},t)\partial_t\psi(\textbf{x},t)-\psi(\textbf{x},t)\partial_t\psi^\ast(\textbf{x},t)\right)=1\,.
\end{eqnarray}

The state of a particle at a given time $t$ may be identified with the ``conditioned'' state $|\psi(t)\rangle:=\sqrt{2\pi}\langle t|\Psi_{m^2}\rangle$, with $|t\rangle=|x^0\rangle$ for $x^0=t$, and thus $\psi(\textbf{x},t)$ with the Klein Gordon wavefunction $\langle{\bf x}|\psi(t)\rangle$. 
In the case of massive particles (positive $m$),  the normalization $\langle \Psi_{\tilde m}|\Psi_{m}\rangle=\delta(m-\tilde{m})$ can instead  be chosen, in which case 
\begin{eqnarray}\label{eq:norm2}
&&\hspace*{-0.5cm}\langle \Psi_{\tilde m}|\Psi_{m}\rangle=\delta(m-\tilde m)\, \Leftrightarrow \nonumber\\ &&\hspace*{-0.5cm}\frac{i}{2m}\!\int\! d^3x \; \left(\psi^\ast(\textbf{x},t)\partial_t\psi(\textbf{x},t)-\psi(\textbf{x},t)\partial_t\psi^\ast(\textbf{x},t)\right)=1,\;\;\;
\end{eqnarray}
i.e., $\int d^{3}x\, \rho(\textbf{x},t)=1$, with $\rho(\textbf{x},t)$  the usual Klein Gordon density \cite{Gr.90, Sc.97}, which in the non relativistic-limit reduces to the Schr\"odinger one for positive energy solutions. 

More generally, it is now easy to prove the following relations
\begin{eqnarray}
\langle \Phi^{\pm}_{\tilde{m}^2}|\Psi^{\pm}_{m^2}\rangle&=& \pm\delta(\tilde{m}^2-m^2) Q(\varphi,\psi)\,, \label{norms1} \\
\langle \Phi^{\pm}_{\tilde{m}^2}|\Psi^{\mp}_{m^2}\rangle&=&0 \,,\label{norm22}
\end{eqnarray}
where the sign $\pm$ indicates the sign of $p^0$ 
and $\langle \Phi_{\tilde{m}^2}|\Psi_{m^2}\rangle$ can be obviously also expressed as $\frac{1}{2\pi} \int d^{4}x \, \varphi^{\ast}(x)\psi(x)$. It is important to notice that the previous relations provide a positive normalization condition for both signs of $p^0$ since $\langle \Psi^{\pm}_{\tilde{m}^2}|\Psi^{\pm}_{m^2}\rangle= \delta(\tilde{m}^2-m^2)|Q(\psi,\psi)|$. 
The positivity follows from the canonical product in $L^2(\mathbb{R}^4)$, yet it implies the usual ``norm''.
The connection between both products can also be derived from extended relations satisfied by the current density. These relations are obtained in the Appendix \ref{A} using the present formalism. The results of Eqs.\ (\ref{norms1}, \ref{norm22}) agree with the general treatment within the quantization of reparametrization-invariant systems \cite{ma.97} (see Sec.\ \ref{sec:prop} and the Appendix \ref{B}). An analogous result which connects a $4d$ invariant product with the $3d$ Dirac's product also holds for Dirac's particles \cite{di.19}.

\subsection{Relationship with Wigner representation}\label{sec:wigner}
The relation between the four and three-dimensional products provides a connection between a fixed mass solution of (\ref{eqgd}) and the usual (scalar) single particle representation in  $L^2(\mathbb{R}^3,d\mu(p))$ where $d\mu(p)=\frac{1}{(2\pi)^3}\frac{d^{3}p}{2 E_{\textbf{p}}}$. The usual improper momentum eigenstates $|\mathbf{p}\rangle_{_{W}} \in L^2(\mathbb{R}^3,d\mu(p))$ are normalized as $_{_{W}}\langle \textbf{p}'|\textbf{p}\rangle_{_{W}}=(2\pi)^3 2 E_{\textbf{p}}\delta^{(3)}(\mathbf{p}-\mathbf{p}')$. We notice that the standard invariant normalization requires the addition of the factor $2E_{\textbf{p}}$ in order to compensate the non invariance of the space volume \cite{M.05,p.18}. 

The connection with the present formalism becomes apparent if we expand 
a solution (\ref{eq:solutions}) 
as, setting   $a(\textbf{p})=\frac{\alpha(\textbf{p})}{\sqrt{(2\pi)^3}}$, $b(\textbf{p})=\frac{\beta(\textbf{p})}{\sqrt{(2\pi)^3}}$,  $E_{\textbf{p}m}\rightarrow E_{\textbf{p}}$  and noting that 
$\delta(p^\mu p_\mu-m^2)H^{\pm}(p^0)=\delta(p^0\mp E_{{\bf p} m})/2E_{\bf p}$,
\begin{eqnarray}
|\Psi_{m^2}\rangle 
&=&\int \frac{d^{3}p}{(2\pi)^3 2 E_{\textbf{p}}} a(\textbf{p})|E_{\textbf{p}m} \textbf{p}\rangle\,,\label{sol21}\\
&&\oplus\int \frac{d^{3}p}{(2\pi)^3 2 E_{\textbf{p}}} b(\textbf{p})|-E_{\textbf{p}m} \textbf{p}\rangle\,,\label{sol22}
\end{eqnarray}
where we have introduced the states
\begin{eqnarray}\label{eq:ep}
|\pm E_{\textbf{p}m}\textbf{p}\rangle:=(2 \pi)^{3/2}\int dp_0\, \delta(p_0\mp E_{\textbf{p}m}) |p_0 \textbf{p}\rangle\,,
\end{eqnarray}
which satisfy ($r,r'=\pm 1$)
) \begin{equation}
\langle r E_{\textbf{p}'m'} \textbf{p}'|r'E_{\textbf{p}m} \textbf{p}\rangle=(2\pi)^3 2 E_{\textbf{p}}\delta_{rr'} \delta^{(3)}(\textbf{p}-\textbf{p}')\delta (m^2-m'^2)\,.
\end{equation}
The factor $2E_{\textbf{p}}$ now arises naturally from the mass orthogonality condition.

The one-to-one correspondence between the states  
\begin{eqnarray} \label{eq:wigner}
|\Psi_{m^2}\rangle&=&\int \frac{d^{3}p}{(2\pi)^{3} 2 E_{\textbf{p}}} a(\textbf{p})|E_{\textbf{p}m} \textbf{p}\rangle\,\in \mathcal{H}\,, \label{eq:Psihistory}\end{eqnarray}
 and the states 
\begin{eqnarray}
|\psi\rangle_{_{W}}&=&\int \frac{d^{3}p}{(2\pi)^{3} 2 E_{\textbf{p}}} a(\textbf{p})| \textbf{p}\rangle_{_{W}}\,\in L^2(\mathbb{R}^3,d\mu(p))\,,\label{eq:psiinst}
\end{eqnarray}
is now explicit since in both cases 
\begin{equation}
    \int \frac{d^3p}{(2\pi)^3 2E_{\bm{p}}} |a(\bm{p})|^2=1,\qquad 
\end{equation}
and their transformation properties are identical. It shall be noticed that while $|\psi\rangle_{_W}$  (Eq. \ref{eq:psiinst}) represents a particle at a fixed time (or equivalently, in the Heisenberg picture), $|\Psi_{m^2}\rangle$ (Eq. \ref{eq:Psihistory}) represents instead the whole history of the particle. In fact, we may also express \eqref{eq:Psihistory} as $|\Psi_{m^2}\rangle=\frac{1}{\sqrt{2\pi}}\int dt\int \frac{d^3p}{(2\pi)^{3}2 E_{\textbf{p}}}e^{-i E_{\textbf{p}}t}a(\textbf{p})|t\textbf{p}\rangle$, 
where $|t\textbf{p}\rangle=\frac{1}{\sqrt{2\pi}}\int dp_0\, e^{ip_0 t}|p_0\textbf{p}\rangle$ 
(notice that $|\textbf{p}\rangle$ differs from $|\textbf{p}\rangle_{_{W}}$) hence defining the proper history state of $|\psi \rangle_{_W}$ in the relativistic framework. 

\subsection{Klein Gordon Propagator}\label{sec:prop}
Given a general state in $\mathcal{H}$, it can be projected onto the subspace of states satisfying (\ref{eqgd}) with a fixed eigenvalue $m^2$ by the operator 
\begin{eqnarray}\label{eq:pi}
\Pi_{m^2}:&=& \delta(\mathcal{J}-m^2)\,.
\end{eqnarray}
In general, this leaves both positive and negative $p^0$ contributions. For the present discussion it is useful to introduce additional projectors $P^{\pm}:=\int dp_0\, H^{\pm}(p_0) |p_0\rangle \langle p_0| \otimes \mathbbm{1}$, satisfying $[P^{\pm},\Pi_m]$=0, and define $\Pi_{m^2}^{\pm}:=P^{\pm}\Pi_{m^2}$.
In particular it is interesting to project $|x\rangle$ onto  the space of ``physical'' particle states:
\begin{eqnarray}\label{eq:proj}
\sqrt{2\pi}\,\Pi^+_{m^2}\, |x\rangle&=& \sqrt{2\pi}\,P^+\delta(\mathcal{J}-m^2)\int \frac{d^{4}p}{\sqrt{(2\pi)^4}}\,e^{ipx} |p\rangle \nonumber \\
&=&  \int \frac{d^{4}p}{\sqrt{(2\pi)^3}}\,\delta(p^\mu p_\mu-m^2) H^+(p_0)e^{ipx} |p\rangle \nonumber \\
&=& \int \frac{d^{3}p}{(2\pi)^3 2 E_{\textbf{p}}}\, e^{i(E_{\textbf{p}}t-\textbf{p}\textbf{x})} |E_{\textbf{p}m} \textbf{p}\rangle\,,
\end{eqnarray}
where the factor $\sqrt{2\pi}$ in the first line was included for normalization (see Eq.\ (\ref{ov4})). 
These states correspond (in the sense discussed in Sec.\ \ref{sec:wigner}) to the single particle states
$\phi(x)|0\rangle$, where $\phi(x)=\int \frac{d^3p}{{(2\pi)^3 \sqrt{2 E_\textbf{p}}}}\left(e^{-ipx} a_{\textbf{p}}+e^{ipx} a^{\dag}_{\textbf{p}} \right)\rvert_{p^0=E_{\textbf{p}}}$ is the Klein Gordon field in the Heisenberg picture for the free theory with mass $m$, and $\sqrt{2 E_{\textbf{p}}}\,a^{\dag}_{\textbf{p}}|0\rangle=|\textbf{p}\rangle_{_W}$. Moreover, from (\ref{eq:proj}) the following identity 
\begin{eqnarray}\label{eq:prop}
2\pi\langle y|\Pi^+_{m^2}\, |x\rangle = \, \langle 0|\phi(y)\phi(x)|0\rangle=D(y-x)\,,
\end{eqnarray}
where 
\begin{equation}
D(y-x)=\int \frac{d^3p}{(2\pi)^32E_{\textbf{p}}} e^{ip(x-y)}|_{p^0=E_{\textbf{p}}}\,,
    \end{equation}
is the \emph{Klein Gordon propagator} (or amplitude) \cite{p.18} for the free theory with mass 
$m$, can be immediately shown.  This expression admits a straightforward interpretation: by selecting the fixed mass  contributions of an event $x$ (see also Sec.\ \ref{sec:normtime}), we obtain a state whose probability to be in another event $y$ is essentially equal to the amplitude for the particle to propagate from $x$ to $y$. We notice that no unitary evolution was explicitly introduced since the states contain all time information. Instead, a proper ``selection'' between possible histories was performed by employing the projector.

\begin{figure}[ht]
    \centering
    \includegraphics[width=3.4 in]{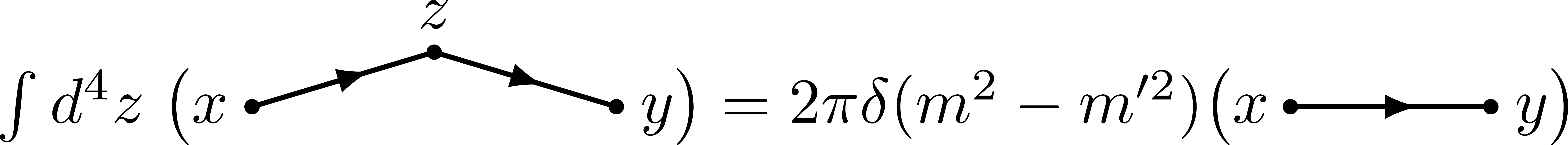}
    \caption{Pictorial representation of the two equivalent characterizations of the quantity $\langle y|\Pi_{m'}^+\Pi_m^+|x\rangle$.
    Each line represents an amplitude $D(y-x)$.}
    \label{fig:feynman}
\end{figure}

From Eq.\ (\ref{eq:prop}) we see that we can rewrite the projection of an event as
\begin{equation}
    \sqrt{2\pi}\Pi^+_{m^2}|x\rangle=\frac{1}{\sqrt{2\pi}}\int d^4z\,D_{m^2}(z-x)|z\rangle\,,
\end{equation}
where we added the index $m^2$ to make the mass dependence explicit. We may also compute the overlap between two projected events as
\begin{eqnarray}\label{eq:proj2}
   2\pi \langle y|\Pi_{{m'}^2}^+\Pi_{m^2}^+|x\rangle&=&2\pi\,\int d^4z\,  \langle y|\Pi_{{m'}^2}^+|z\rangle\langle z|\Pi_{m^2}^+|x\rangle \nonumber \\
       &=&2\pi\delta(m^2-{m'}^2)\, \langle y|\Pi_{m^2}^+|x\rangle   \label{ov3}\\
       &=&\delta(m^2-{m'}^2)D(y-x)
   \label{ov4}
\end{eqnarray}
 where in (\ref{ov3}) we have employed Eq.\ (\ref{eq:proj}). Thus,  with the normalization employed  for the projected events their overlap is directly the propagator times the mass delta function.  
 The identity \eqref{ov3} implies 
 \begin{equation}
 \int d^4z\,D_{{m'}^2}(y-z) D_{m^2}(z-x)=2\pi\delta(m^2-{m'}^2)D(y-x)\,.
\end{equation}
 The finite part is again essentially the propagator 
  while the presence of the delta function is in agreement 
 with  the discussion of Sec.\ \ref{sec:scalarp}. However, we see from Eq.\ (\ref{eq:proj2}) that we can reinterpret the appearance of the Dirac delta as the result of 
 summing over all possible space-time points $z$ in the propagation from
 $x$ to $y$ with the additional intermediate point $z$. This result is pictorially represented in Fig.\ \ref{fig:feynman}. 
 
In group averaging techniques the result (\ref{eq:proj2}) is employed to induce the inner product of the physical Hilbert space \cite{ma.00} which in this case corresponds to a particle with fixed mass. In the present notation this can be stated as follows: Let $|\Phi_{m^2}\rangle:=\Pi_{m^2}|\Phi\rangle$ and $|\Psi_{m^2}\rangle$ be two solutions of the constraint (\ref{eqgd}), then 
     $(\Phi_{m^2}|\Psi_{m^2})_{\text{phys}}:=\langle \Phi|\Psi_{m^2}\rangle$, 
which is equivalent to the relations (\ref{norms1}, \ref{norm22}) without the Dirac delta in (\ref{norms1}). In the present approach we preserve the mass delta since the extended Hilbert space is considered physically relevant as pointed out in the following sections. As a consequence, the ``physical'' subspaces of $\mathcal{H}$ are genuine subspaces (the space of solutions of (\ref{eqgd}) and $\mathcal{H}$ share the same inner product).

We also mention that $\Pi_{m^2}$ has the formal representation $\Pi_{m^2}=\frac{1}{2\pi}\int_{-\infty}^{\infty} d\tau\,\exp[i\tau(\mathcal{J}-m^2)]$, resembling proper time methods \cite{sc.51}. In fact, the result of restricting the same integral to positive $\tau$ (and adding an infinitesimal imaginary part $i \epsilon$) is proportional to the inverse  operator of $\mathcal{J}-m^2$, whose matrix elements are equal to the Feynman propagator and for which an asymptotic projective meaning holds \cite{ma.97}.

\subsection{Normalization in Time}\label{sec:normtime}
A state of the form 
\begin{equation} \label{eq:rep}
|\Psi\rangle= \int dm^2 (\gamma^+\phi^+(m^2)|\Psi^+_{m^2}\rangle+\gamma^-\phi^-(m^2)|\Psi^-_{m^2} \rangle)\,, 
\end{equation}
where $|\Psi^\pm_{m^2}\rangle$ are normalized states defined as in (\ref{norms1}) ($\langle \Psi^\pm_{{m'}^2}|\Psi^\pm_{m^2}\rangle=\delta(m^2-{m'}^2))$ with 
\begin{equation}
 \int dm^2 |\phi^{\pm}(m^2)|^2=1\,,
\end{equation} and
\begin{equation}
\langle \Psi|\Psi\rangle=|\gamma^+|^2+|\gamma^-|^2=1 \,,
\end{equation}
belongs to $L^2(\mathbb{R}^4)$. 
We will now prove that any state $|\Psi\rangle \in L^2(\mathbb{R}^4)$ admits the representation (\ref{eq:rep}). This is in principle apparent as the integral over all real values of $m^2$ covers the spectrum of the hermitian operator ${\mathcal J}$ and $|\Psi^{\pm}_{m^2}\rangle$ are general states with definite mass and sign of $p^0$.  
This also means that consideration of states which are normalizable in time  (e. g. finite time history) is equivalent to allow a mass/$p^0$ sign uncertainty. The states $|\Psi^+_{m^2}\rangle$ may be regarded as the idealization corresponding to a particle with infinite history and infinitely well defined dispersion relation, in which case the correspondence of Sec.\ \ref{sec:wigner} follows.

\textit{Proof.} An arbitrary normalized state $|\Psi\rangle \in L^2(\mathbb{R}^4)$ can be expanded as
\begin{eqnarray}
    |\Psi\rangle&=&\int d^4p
    \langle p|\Psi\rangle|p\rangle=\int\! d^4p \int\! d m^2 \delta(p^\mu p_\mu-m^2)]\langle p|\Psi\rangle |p\rangle\nonumber\\
    &=&\int d m^2\left[\int \frac{d^3p}{(2\pi)^{3}2E_{{\bf p}m}}\langle  E_{{\bf p},m}{\bf p}|\Psi\rangle | E_{{\bf p}m}\bf{p}\rangle\right.\label{p0p}\\
    &&+\left.\int \frac{d^3p}{(2\pi)^{3}2E_{{\bf p}m}}\langle -E_{{\bf p},m}{\bf p}|\Psi\rangle |-E_{{\bf p}m}\bf{p}\rangle\right]\label{p0m}
\end{eqnarray}
where $\int dm^2\ldots=\int_{0}^\infty dm^2\ldots+\int_{-\infty}^0 dm^2\ldots$
includes all real values of $m^2$. Using Eqs.\ \eqref{sol21}--\eqref{sol22}, Eqs.\ (\ref{p0p})--(\ref{p0m}) are seen to be of the form (\ref{eq:rep}) with $a({\bf p})=\langle  E_{{\bf p},m}{\bf p}|\Psi\rangle/(\gamma^+\phi^+(m^2))$, $b({\bf p})=\langle -E_{{\bf p},m}{\bf p}|\Psi\rangle/(\gamma^-\phi^-(m^2))$ and 
\[\gamma^{\pm}\phi^{\pm}(m^2)=
\sqrt{\int \frac{d^3p}{(2\pi)^3 2E_{{\bf p}m}}
|\langle\pm E_{{\bf p}m}{\bf{p}}|\Psi\rangle|^2}\,.\]
They involve four distinct terms, according to the signs of $m^2$ and $E_{{\bf p}m}$. For $m^2<0$ the $d^3p$ integration is restricted to the region $|{\bf p}|^2>-m^2$, as depicted in Fig.\ 2.  \qed

The four terms which arise from decomposing a general state $|\Psi\rangle\in L^2(\mathbb{R}^4)$ according to the signs of $m^2$ and $p^0$ in Eqs.\ (\ref{p0p})--(\ref{p0m}) belong to orthogonal subspaces which are Hilbert space representations of the corresponding classes of irreducible representations of the Poincar\'e group \cite{w.39, Bog.75}. This exhaustivity of $\mathcal{H}$ is precisely what allows to represent events $|x\rangle$ and in particular the definition of a time operator $T:=X^0$ such that $X^0|x\rangle=x^0|x\rangle$. The time translation operator $P^0=\int d^4p\, p^0 |p\rangle \langle p|$ is, as expected, unbounded, however, this is not a problem in the present formalism, in contrast with other approaches \cite{U.89, qg}: By writing (as in Eqs.\ (\ref{p0p})--(\ref{p0m}))
\begin{eqnarray}
    P^0&=&\int d m^2\left[\int \frac{d^3p}{(2\pi)^{3}2E_{{\bf p}m}}E_{{\bf p}m}|E_{{\bf p}m}\textbf{p}\rangle\langle  E_{{\bf p}m}\textbf{p}|\right.\\
    &&-\left.\int \frac{d^3p}{(2\pi)^{3}2E_{{\bf p}m}}E_{{\bf p}m} |-E_{{\bf p}m}\textbf{p}\rangle\langle -E_{{\bf p}m}\textbf{p}|\right]\,,
\end{eqnarray}
it becomes clear that all four regions of $\mathcal{H}$ contribute to its spectrum leading, as a consequence, to its unboundedness.
Instead, on states which belong to a particular irreducible representation, imposed `a posteriori' by Eq.\ (\ref{eqgd}) and by a given choice of the sign of $p^0$, $P^0$ acts properly:
\begin{equation}
    P^0|\pm E_{{\bf p}m}\textbf{p}\rangle=\pm E_{{\bf p}m}|\pm E_{{\bf p}m}\textbf{p}\rangle\,.
\end{equation}
The advantage of the present approach is 
 apparent: covariant operators are defined independently of the particular theory, still, after a given theory, or superposition of them, is chosen, these operators, which are still defined as before, act properly. This is precisely what we have already found by projecting an event in Sec\ \ref{sec:wigner}: the state $|x\rangle$, to which we associate a geometrical meaning, is ``unphysical'' for a theory with fixed mass $m^2$ and $p^0>0$, however, the ``closest'' physical state corresponds to the well known state $\phi(x)|0\rangle$. Moreover, a perturbative treatment of an interacting theory implicitly involves states with an undefined mass when expanded in terms of the free basis. This can already be discussed within a ``first quantization'' treatment of interactions as shown in Sec.\ \ref{sec:ext}.

\begin{figure}[ht]
    \centering
    \includegraphics[width=0.35\textwidth]{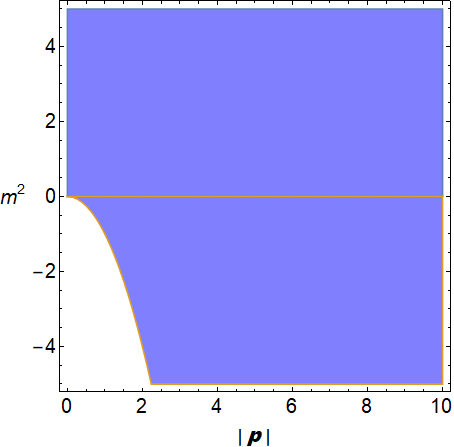}
    \caption{Integration region in the variables $m^2$ and $\textbf{p}$. Here $|\textbf{p}|$ is the modulus of the three-momentum $\textbf{p}$.}
    \label{fig:region}
\end{figure}

\section{Klein Gordon equation in an external field}\label{sec:ext}
So far the discussion was centered on the case of a free particle. In this section we discuss interactions at first quantization level by treating fields as external entities. This will follow from a straightforward extension of the previous ideas which, remarkably,  still provides the right connection between the invariant norm and the Klein Gordon normalization, and more generally, between the canonical product in $L^2(\mathbbm{R}^4)$ and the Klein Gordon product. 
We replace $\mathcal{J}=P^{\mu}P_{\mu}$ by
\begin{equation}
\mathcal{J}_A=(P^{\mu}+eA^{\mu}(X))(P_\mu+eA_\mu(X))\,, \label{JA}   
\end{equation}
with $A_{\mu}(X)|x\rangle=A_{\mu}(x)|x\rangle$. A state $|\Psi\rangle=\frac{1}{\sqrt{2\pi}}\int d^{4}x \,\psi(x) |x\rangle$ satisfies 
\begin{equation}\label{eq:kga}
    \mathcal{J}_A |\Psi\rangle=m^2|\Psi\rangle\,,
\end{equation}
iff $\psi(x)$ satisfies the Klein Gordon equation
\begin{eqnarray}\label{eq:kgx}
\left((-i\partial_\mu+eA_{\mu})(-i\partial^{\mu}+e A^{\mu})-m^2\right)\psi(x)=0\,.
\end{eqnarray}

Let us now consider the case where $A_\mu(X)$ does not depend on $T$ nor $m^2$. Considering solutions of definite energy $E$, $\psi(x)=e^{-iEt} \psi({\bf x})$, Eq.\ \eqref{eq:kgx} leads to the equation 
\begin{equation}\label{eq:FE}
    [F(E)-m^2]\psi({\bf x})=0\,,
\end{equation}
where $F(E)=(E-eA_0)^2-({\bf P}+e{\bf A})^2$ is an hermitian operator with respect to the standard product in $L^2(\mathbb{R}^3)$, which does not depend on $m^2$. 
Then we may write a general solution of \eqref{eq:kgx} as 
\begin{eqnarray}\psi_{m^2}(x)&=&\sum_k c_k\psi_k(x,m^2),\\
\psi_k(x,m^2)&=&e^{-iE_k(m^2) t}\psi_k({\bf x},m^2)\,,\label{psikm0}\end{eqnarray} 
where $k$ labels the modes of definite energy $E_k(m^2)$ and  $\psi_k({\bf x},m^2)$ the corresponding eigenfunctions, obtained from Eq.\ \eqref{eq:FE}. 
They satisfy the Klein Gordon orthogonality $Q_A(\psi_{k'}(m^2),\psi_{k}(m^2))=0$ for $E_{k}(m^2)\neq E_{k'}(m^2)$, 
where  ($D_\mu=\partial_\mu-ieA_\mu({\bf x})$) 
\begin{eqnarray}
    Q_{A}(\phi,\psi)&=&
    i\int d^3x\,\left(\phi^*({\bf x},t)D_0\psi({\bf x},t)-
    \psi({\bf x},t)D_0^*
    \phi^*({\bf x},t)\right)\,.\nonumber\\&&\label{QA}
\end{eqnarray}
The ensuing solution of \eqref{eq:kga} is  
\begin{eqnarray}\label{eq:solA}
    |\Psi_{m^2}\rangle&=&\sum_k c_k|\Psi_{k}(m^2)\rangle\,,\\
|\Psi_{k}(m^2)\rangle&=&
\frac{1}{\sqrt{2\pi}}\int d^4x\, \psi_k(x,m^2)|x\rangle\,.\label{psikm}\end{eqnarray}

We will prove in the first place that solutions 
with definite energies $E_k(m^2)$ 
satisfy the orthogonality condition 
\begin{equation}\label{eq:orten}
\langle \Psi_{k'}({m'}^2)|\Psi_{k}(m^2)\rangle=\delta(m^2-{m'}^2)
Q_{A}(\psi_{k'},\psi_k)s_{k}\end{equation}
where $s_{k}=\text{sgn}\left(\frac{dE_k}{dm^2}\right)$ and the left hand side is the canonical product in $L^2(\mathbb{R}^4)$. 
This is a non trivial result which follows from ``special'' orthogonality relations of the usual solutions of Klein Gordon equation, as shown below. While it warrants the expected orthogonality of eigenstates  with different masses, at equal mass it directly links the standard product in $\mathbb{R}^4$ with the Klein Gordon product in $\mathbb{R}^3$, which in turn ensures orthogonality of states with different energies at equal mass and implies $Q_{A}(\psi_{k},\psi_k)s_{k}=|Q_{A}(\psi_{k},\psi_k)|$.  

Secondly, we will show, choosing orthogonal modes $\psi_k(m^2)$ ($Q_A(\psi_{k'}(m^2),\psi_k(m^2))=0$ for $k\neq k'$,  that relation \eqref{eq:orten}  implies
\begin{equation}
    \langle \Psi'_{m'^2}|\Psi_{m^2}\rangle=\delta(m'^2-m^2)\sum_k {c'_k}^*c_k|Q_A(\psi_k,\psi_k)|\,,\label{gral}
\end{equation}
for general state with definite mass, which is identical with $\delta(m'^2-m^2)|Q_A(\psi',\psi)|$ when all $Q_A(\psi_k,\psi_k)$ have the same sign (i.e., all ``positive'' energy modes in standard conditions). This is the sought extension of Eqs.\ \eqref{norms1}. 

\textit{Proof.} The overlap between two solutions (\ref{psikm})  with definite energies yields
\begin{eqnarray}\label{overlapproof}
    \langle \Psi_{k'}(m'^2)|\Psi_{k}(m^2)\rangle=\delta(E_{k'}(m'^2)-E_{k}(m^2))\nonumber\\\times \int d^3x\,\psi^\ast_{k'}(\textbf{x},m'^2)\psi_k(\textbf{x},m^2)\,.
\end{eqnarray}
 States with different energies are automatically orthogonal while the equal energies condition can be separated into two contributions:
 equal energies at equal masses, or equal energies at different masses (and different $k$). Consider first the second case: by writing 
 \begin{eqnarray}
     \left[F(E_k(m^2))-m^2\right]\psi_{k}(\textbf{x},m^2)&=&0 \label{eq:solproof}\\
      \left[F(E_{k'}(m'^2))-m'^2\right]\psi_{k'}(\textbf{x},m'^2)&=&0\,,
 \end{eqnarray}
multiplying on the left by $\psi^\ast_{k'}(\textbf{x},m'^2)$ ($\psi^\ast_{k}(\textbf{x},m^2)$) the first (second) equation, integrating in the whole space and subtracting (conjugating one of the results) we obtain
\begin{align}
    (m'^2-m^2)\int d^3x\, \psi^\ast_{k'}(\textbf{x},m'^2)\psi_{k}(\textbf{x},m^2)=\nonumber\\ 
     (E_k(m^2)-E_{k'}({m'}^2))Q_A(\psi_{k'}({m'}^2),\psi_{k}(m^2))&&\label{mpm}
   \end{align}
 where we have used the hermiticity of $({\bf P}+e{\bf A})^2$.  For $E_{k'}(m'^2)=E_k(m^2)$ then
\begin{equation}
      (m'^2-m^2)\int d^3x\, \psi^\ast_{k'}(\textbf{x},m'^2)\psi_{k}(\textbf{x},m^2)=0\,,\label{eq:mmp}
\end{equation}
implying  an extended orthogonality condition for $m'^2\neq m^2$ when energies coincide.  We conclude that no contributions from different masses actually arise in \eqref{overlapproof}. Note also that for ${m'}^2=m^2$ but $E_{k}(m^2)\neq E_{k'}(m^2)$ Eq.\ (\ref{mpm}) leads to the 
standard Klein Gordon orthogonality condition $Q_A(\psi_{k'}(m^2),\psi_{k}(m^2))=0$. 

Previous results \eqref{overlapproof}, \eqref{eq:mmp} allows us to write, for modes of equal 
energies  ($E_k(m^2)=E_{k'}(m^2) \,\forall m^2$) 
\begin{eqnarray}\label{eq:overlapproof2}
   \langle \Psi_{k'}(m'^2)|\Psi_{k}(m^2)\rangle=\frac{\delta(m'^2-m^2)}{|dE_k/dm^2|}\nonumber\\
   \times \int d^3x\,\psi^\ast_{k'}(\textbf{x},m^2)\psi_k(\textbf{x},m^2)\,.
\end{eqnarray}
This second part of the proof involves finding an expression for $dE_k/dm^2$. This is achieved by deriving Eq. (\ref{eq:solproof}) with respect to $m^2$, which yields
    \[[F'(E_k)\frac{dE_k}{dm^2}-1]\psi_k(\textbf{x},m^2)=\left[F(E_k)-m^2\right]\frac{d\psi_k(\textbf{x},m^2)}{dm^2}\,.\]
with $F'(E_k)=2(E_k-eA_0)$. We now multiply on the left by a solution with the same energy $\psi^\ast_{k'}(\textbf{x},m^2)$ and integrate in space; we obtain
\begin{equation}
    \int d^3x\, \psi^\ast_{k'}(\textbf{x},m^2)\psi_k(\textbf{x}, m^2) \left[F'(E_k)\frac{dE_k}{dm^2}-1\right]=0\,,\label{dF}
\end{equation}
and thus, for $E_k(m^2)=E_{k'}(m^2)$, 
\begin{equation}
    \int d^3x\, \psi^\ast_{k'}(\textbf{x},m^2)\psi_k(\textbf{x}, m^2)=
    \frac{dE_k}{dm^2}Q_A(\psi_{k'}(m^2),\psi_k(m^2))
\end{equation}
which is the natural extension extension of \eqref{mpm} for $m^2={m'}^2$ and $E_k(m^2)=E_{k'}(m^2)$. 
Inserting this relation in \eqref{eq:overlapproof2} leads to the result \eqref{eq:orten}. 
Eq.\ (\ref{dF}) also reveals an additional orthogonality condition: orthogonal modes at equal energies according to Klein Gordon product are also orthogonal in the canonical product of $L^2(\mathbb{R}^3)$, assuming $\frac{dE_k}{dm^2}\neq 0$. \qed

Finally,  we note from Eq.\ \eqref{eq:orten} that imposing the normalization  
$\langle \Psi_{k'}({m'}^2)|\Psi_{k}(m^2)\rangle=\delta(m^2-{m'}^2)\delta_{kk'}$ directly leads to the Klein Gordon normalization 
$|Q_{A}(\psi_{k'}(m^2),\psi_k(m^2))|=\delta_{kk'}$.
 
The rigorous extension of the present results to a general potential $A_\mu(X)$ and curved space-times involves new concepts and will be presented elsewhere. Nevertheless, 
general identities for the current density in the presence of a general potential are discussed in the Appendix \ref{A}. 
The case of a mass dependent $A_\mu$ is briefly discussed in the non relativistic limit for Newtonian gravity in Sec.\ \ref{sec:nonrel}. 

The results of this section can be directly employed to define a physical Hilbert space at fixed $m^2$ (as mentioned before for the free case) 
replacing $\Pi_{m^2}\to \delta(\mathcal{J}_A-m^2)$, extending then previous quantization programs \cite{ma.97, ma.00} to the case where an external $A_\mu$ is present. However, and maybe more importantly, we observe that the mass eigenstates of ${\cal J}_A$ in Eq.\ (\ref{eq:kga}) are obviously not eigenstates of the free particle ${\cal J}$ of Eq. \eqref{eqgd}, since ${\cal J}_A$ and ${\cal J}$ do not commute. Therefore, the expansion of eigenstates $|\Psi_{m^2}^A\rangle$ of ${\cal J}_A$ in terms of those of ${\cal J}$ generally involves an expansion over different masses (and may also involve negative energies) as that considered in Sec.\ \ref{sec:normtime}. Thus, the consideration of states with no definite mass in the free basis representation is already implicit when dealing with an external field, i.e., with interactions. This in turn
reveals that the extended Hilbert space, commonly considered as an auxiliary construction, plays an unavoidable physical role in a $4d$ formalism. Besides, any fluctuation of the fields $A_\mu$, which in a more realistic scenario are also dynamical, would lead the system to explore different mass sectors of $\mathcal{H}$. 

\section{Non Relativistic Limit}\label{sec:nonrel}
It is well known that for positive energy solutions in the non relativistic limit $E'/m \ll 1$ (order $ (v/c)^2$, with $E'=E-m$) the Klein Gordon equation reduces to the Schr\"odinger equation \cite{Gr.90}. In particular the Klein Gordon norm for massive particles becomes the standard Schr\"odinger norm. 
It is then to be expected that a non relativistic version of Eq.\ (\ref{eq:norm2}) in terms of the usual quantum mechanical norm holds as a limit. Indeed this is the case, but it is instructive to derive this result directly from the non relativistic regime. 

We first recall that Schr\"odinger equation can be recovered for states $|\psi(t)\rangle \in \mathcal{H}_S$ by imposing a global static constraint on states  $|\Psi\rangle\in \mathcal{H}=\mathcal{H}_T\otimes \mathcal{H}_S$. Here $\mathcal{H}_T$ is spanned by the eigenstates $|t\rangle$ of the operator $T$ which satisfies the canonical commutation $[T,P_{T}]=i$. In PaW interpretation \cite{PaW.83} $\mathcal{H}_T$ is regarded as the Hilbert space of a quantum clock such that the parameter $t$ is a label of states $|t\rangle$ of this particular system.

The states $|\Psi\rangle$ can be expanded as 
\begin{equation}\label{eq:clocksystemstate}
|\Psi\rangle=\int dt\, |t\rangle |\psi(t)\rangle\,,
\end{equation}
while the state of the system at ``time'' $t$ is $|\psi(t)\rangle=\langle t | \Psi \rangle$. 
By imposing the equation
\begin{equation}\label{eq:wdwnr}
\mathcal{J}|\Psi\rangle=0\,,
\end{equation}
with 
\begin{equation}\label{eq:wdwoperator}
\mathcal{J}=P_{T}\otimes \mathbbm{1}+\mathbbm{1}\otimes H\,,
\end{equation}
where $H$ is the Hamiltonian of the system, the standard Schr\"odinger equation is obtained \cite{QT.15}:
\begin{equation}\label{eq:implic}
\langle t|\mathcal{J}|\Psi \rangle=0\Rightarrow i\frac{d}{dt}|\psi(t)\rangle=H|\psi(t)\rangle\,.
\end{equation}

  The previous implication holds for arbitrary Hamiltonians iff the spectrum of $P_T$ is the entire real line, which also implies the same spectrum for $T$. Under this condition the states $|\Psi\rangle$ cannot be normalized in $\mathcal{H}$ \cite{QT.15}: roughly speaking, $ \langle \Psi|\Psi\rangle$ is equal to $\langle\psi(t_0)|\psi(t_0)\rangle$ times the (infinite) length of time. 
  On the other hand, if we focus on the case of a scalar particle, from the discussion of Sec.\ \ref{sec:normtime}  we can infer how to properly relate the norm of these global states with the norm of $|\psi(t_0)\rangle$. It is also important to notice that in the relativistic case the quantity $\langle \Psi|\Psi\rangle$ is not related to the length of time as before since the conditioned states $|\psi(t)\rangle$ are normalized according to the Klein Gordon norm, not the Schr\"odinger one.
     We now focus then on the case $\mathcal{H}_S=\text{span}\{|\textbf{x}\rangle\}$
   so that
   \begin{equation}
       |\Psi\rangle=\int dtd^3x\, |t\rangle|\textbf{x}\rangle \langle \textbf{x}|\psi(t)\rangle\equiv\int dtd^3x\, \psi(\textbf{x},t)|t\textbf{x}\rangle\,,
   \end{equation}
and consider first the free case $H=\frac{P^2}{2m}$. Notice that for the quantum mechanical point of view, the zero eigenvalue in Eq.\ (\ref{eq:wdwnr}) plays no special role since a shifted eigenvalue of $\mathcal{J}$ corresponds to a global energy translation. On the other hand, since we are dealing with a free particle it is wise to set the rest energy to the (positive) value $mc^2$ (where we have momentarily reintroduced the speed of light $c$). Then we have
\begin{equation}\label{eq:wdwm}
   \langle t \textbf{x}| (\mathcal{J}+m)|\Psi\rangle=0\Rightarrow \left(-i\frac{\partial}{\partial t}+ \frac{-\nabla^2 }{2m}+m \right)\psi(\textbf{x},t)=0\,,
\end{equation}
whose solutions are clearly of the form
\begin{equation}
 |\Psi_m\rangle=\int d^3p\, a(\textbf{p}) |p^2/2m+m, \textbf{p}\rangle\,,
\end{equation}
implying
\begin{eqnarray}
    \langle \Psi_{m'}|\Psi_m\rangle&\approx&\delta(m-m')\int d^3p \frac{|a(\textbf{p})|^2}{|1-\frac{p^2}{2m^2}|}\nonumber \\
    &=&\delta(m-m')\int d^3p\, |a(\textbf{p})|^2\times\left(1+O\left(\frac{E'_p}{m}\right)\right)\nonumber
\end{eqnarray}
(we assume  $|a(\textbf{p})|$ significant only for $p\ll m,m'$).  Up to $O(p^2/m^2)$, this equation coincides with Eq.\ (\ref{eq:deltanorm}) after replacing $a({\textbf p})=\alpha({\bf p})/(2E_{{\bf p}})$.  
We find that in this regime solutions with different ``eigenvalues'' $m$ are orthogonal 
, even if $\mathcal{J}$ is mass-dependent. Moreover, for states of non relativistic momenta, which is precisely the regime of validity of Schr\"odinger equation for a particle, the orthonormality condition implies the Schr\"odinger norm (up to relativistic corrections): 
\begin{equation}\label{eq:normsch}
\langle \Psi_{m'}|\Psi_m\rangle=\delta(m-m') \Rightarrow \langle \psi(t_0)|\psi(t_0)\rangle=1+O(v^2/c^2)\,.
\end{equation}
This also means that the history states $|\Psi\rangle$ can be normalized according to the discussion of Sec.\ \ref{sec:normtime}, a result which provides (in the present case) a physical interpretation to the regularization proposal of \cite{QT.15}.

This result can be easily extended in this same context (but it also follows from the non relativistic limit of Sec.\ \ref{sec:ext}) to Hamiltonians of the form 
\begin{equation}
    H=\frac{(\textbf{p}-e\textbf{A}(\textbf{x}))^2}{2m}+V(\textbf{x})+m\phi(\textbf{x})+m\,,
\end{equation}
where $\textbf{A}$, $V$ and $\phi$ are mass independent, employing a similar strategy of Sec.\ \ref{sec:ext} which was already employed for Dirac Hamiltonian in \cite{di.19}. 
A minor modification follows from the mass dependent potential $m \phi(\textbf{x})$: since now $\langle x|(P_T+H)|\Psi_m\rangle=0$ yields 

\begin{eqnarray}
    \left(i\frac{\partial}{\partial t}- \frac{(\textbf{p}-e\textbf{A}(\textbf{x}))^2 }{2m}-V(\textbf{x})\right)\psi(\textbf{x},t)&&\nonumber \\
    =m(1+\phi(\textbf{x}))\psi(\textbf{x},t)&&\,,
\end{eqnarray}
 the universe equation must be considered as a generalized eigenfunction equation (up to relativistic corrections coming from the mass dependent term on the left hand side). In order to achieve orthogonality the generalized product $(\Psi|\Psi):=\langle \Psi|(1+\phi(\textbf{X}))|\Psi\rangle=\int d^4x\, (1+\phi({\textbf{x}}))  |\psi(\textbf{x},t)|^2$
 must in principle be employed. However, if $c$ is reintroduced, $(\Psi|\Psi)=\int d^4x\, (1+\phi(\textbf{x})/c^2)|\psi(\textbf{x},t)|^2$ and we see that $\phi(\textbf{x})/c^2$ must be dropped at this order \cite{pa.11}. This implies that potentials which depend on mass linearly, as a Newtonian gravitational potential, do not  require a special treatment at the lowest order in $c$. It's still remarkable that this simple analysis suggests  a connection between gravity and curvature since only Quantum Mechanical and Newtonian gravity considerations were made together with the rest mass energy condition $E=mc^2$.

\section{Extended Fock space}\label{sec:fock}
 In this section we explore the construction of a Fock space $\mathcal{H}_{EF}$ where the building block is the single particle (sp) basis $\{|p\rangle\}$, while the corresponding usual sp in $\mathcal{H}_F$ is $\{a_\textbf{p}^\dag|0\rangle_F=|\textbf{p}\rangle\}$.
 The states $|p\rangle$ are reinterpreted as the basis of a single particle space, i.e. 
\begin{equation}
   |p\rangle =c_p^\dag|0\rangle\,, 
\end{equation}
where the creation/annihilation operators satisfy, since $\langle p'|p\rangle=\delta^4(p-p')$, the algebra
\begin{eqnarray}
    [c_p,c_{p'}^\dag]&=&\delta^{(4)}(p-p')\,, \\
   {[c_p,c_{p'}]}&=&{[c^\dag_p,c^\dag_{p'}]=0}
   \,.
\end{eqnarray}
 This algebra is explicitly preserved by boost operators whose definition,
\begin{equation}\label{eq:boost}
    U(\Lambda) c_p  U^\dag(\Lambda)=c_{\Lambda p}\,,
\end{equation}
 follows from Eq.\ (\ref{ust}). 
 Note that $U(\Lambda)=\exp[-\frac{i}{2}w^{\mu\nu} L_{\mu\nu}]$ is explicitly unitary and that  \[L_{\mu\nu}=i\int d^4p\, c^\dag_p\left(p_\mu\frac{\partial}{\partial p^\nu}-p_\nu\frac{\partial}{\partial p^\mu}\right)c_p\,,\] the generator of Lorentz transformations, is an hermitian one-body operator.  
    
Defining ${\cal J}$ as the one-body operator 
 \begin{equation}
    \mathcal{J}=\int d^4p\; (p^2-m^2)c^\dag_pc_p\,,
\end{equation}
 which is the universe operator that corresponds to (\ref{JD}), 
 the physical subspace is defined by those states built from creation operators commuting with ${\cal J}$: 
\begin{equation}\label{eq:constraint}
    [\mathcal{J},c^\dag_{p}]=(p^2-m^2)c^\dag_p=0\Rightarrow p^2=m^2\,.
    \end{equation}

As a basis of this subspace we can employ, for $p^0>0$, the operators
\begin{equation}
    c^{(m)}_{\textbf{p}}:=\sqrt{\frac{(2\pi)^3}{2 E_{\textbf{p}m}}}\int dp_0\, \delta(p_0-E_{\textbf{p}m})\, c_{p_0 \textbf{p}}\,,\label{cpm}
\end{equation}
which satisfy
\begin{equation}\label{eq:conm}
    [c^{(m)}_{\textbf{p}},c^{\dag(m')}_{\textbf{p}'}]=(2\pi)^3 \delta^{(3)}(\textbf{p}-{\textbf{p}}')\delta(m^2-m'^2)\,,
\end{equation}
and transform, according to (\ref{eq:boost}), as 
\begin{equation}\label{eq:cboost}
    U(\Lambda)c_{\textbf{p}}^{(m)}U^\dag(\Lambda)=\sqrt{\frac{E_{\Lambda \textbf{p}m}}{E_{\textbf{p}m}}}c^{(m)}_{\Lambda\textbf{p}}\,.
\end{equation}

A single particle state of mass $m$ is then written as
\begin{eqnarray}
    |\Psi_{m^2}\rangle&=&\int \frac{d^3p}{(2\pi)^3\sqrt{ 2E_{\textbf{p}m}}}a(\textbf{p})c^{\dag (m)}_{\textbf{p}}|0\rangle\\ \nonumber
    &=&\int \frac{d^3p}{(2\pi)^3 2E_{\textbf{p}m}}a(\textbf{p})|E_{\textbf{p}m}\textbf{p}\rangle\,,
\end{eqnarray}
where
\begin{equation}
    |E_{\textbf{p}m}\textbf{p}\rangle:=\sqrt{ 2E_{\textbf{p}m}}c^{\dag (m)}_{\textbf{p}}|0\rangle\,.
\end{equation}

According to the discussion of Sec.\ \ref{sec:wigner}, the state $|\Psi_{m^2}\rangle$ can be identified with the history of 
\begin{eqnarray}
    |\psi\rangle_{_W}&=&\int \frac{d^3p}{(2\pi)^3\sqrt{ 2E_{\textbf{p}m}}}a(\textbf{p})a^{\dag }_{\textbf{p}}|0\rangle_F\\ \nonumber
   &=&\int \frac{d^3p}{(2\pi)^3 2E_{\textbf{p}m}}a(\textbf{p})|\textbf{p}\rangle_{_W}\,,
\end{eqnarray}
where  $a_{\textbf p}$ are the standard $3d$ operators: 
\begin{eqnarray}\label{eq:conmusual}
    [a_{\textbf{p}},a_{\textbf{p}'}^\dag]=(2\pi)^3\delta^{(3)}(\textbf{p}-\textbf{p}')
\end{eqnarray}
with 
\begin{equation}
    |\textbf{p}\rangle_{_W}=\sqrt{2 E_{\textbf{p}m}}a_{\textbf{p}}^\dag|0\rangle_F\,.
\end{equation}

 It is now straightforward to extend this identification to many particles. 
 From the usual transformation law of the operators $a_\textbf{p}$, $a^\dag_{\textbf{p}}$, and Eq.\ (\ref{eq:cboost}) it follows that these identifications are frame independent.

It is now interesting to consider a two particle state 
 \[   |\Psi\rangle=\int \frac{d^3p_1}{(2\pi)^3\sqrt{2E_{\textbf{p}_1}}}\frac{d^3p_2}{(2\pi)^3\sqrt{2E_{\textbf{p}_2}}}a(\textbf{p}_1,\textbf{p}_2)c^{\dag (m)}_{\textbf{p}_1}c^{\dag (m)}_{\textbf{p}_2}|0\rangle \]
    where $c^{\dag (m)}_{\textbf{p}_i}\propto c_{p^0{\textbf p}}$ with $p^0=E_{{\textbf p}m}$  are the operators defined in (\ref{cpm}). By a Fourier transform in $p^0_1$, $p^0_2$, we obtain 
    \begin{eqnarray}|\Psi\rangle&=&\int \frac{d^3p_1dt_1d^3p_2dt_2\,e^{-iE_{\textbf{p}_1}t_1}e^{-iE_{\textbf{p}_2}t_2}}{(2\pi)^3\sqrt{2\pi}2E_{\textbf{p}_1}(2\pi)^3\sqrt{2\pi}2E_{\textbf{p}_2}}
    \nonumber\\&&\times\, a(\textbf{p}_1,\textbf{p}_2)c^\dag_{t_1\textbf{p}_1}c^\dag_{t_2\textbf{p}_2}|0\rangle \,. \label{eq:twopart}
\end{eqnarray}
It is then seen that this state has not a simple time structure of the form $\int dt d^3p_1d^3p_2\, \psi(t,p_1,p_2)c^\dag_{t\textbf{p}_1}c^\dag_{t\textbf{p}_2}|0\rangle$. This is relevant since such form cannot be preserved by Lorentz transformations. The more complex structure of $|\Psi\rangle$ is a 
novelty of the relativistic case which is required to represent boosts.

\section{Discussion}\label{sec:disc}
We have seen that it is possible to construct a consistent history state formalism for a scalar particle whose concept of time shares the underlying mathematical ideas of the Page and Wooters mechanism \cite{PaW.83}. 
The consideration of a suited Hilbert space for the representation of explicitly covariant operators, together with a timeless universe equation allows a simple derivation of the Klein Gordon equation, both in the free case and with an external field,  that complements the previous derivations of the Schr\"odinger \cite{QT.15} and Dirac equations \cite{di.19}. 
The canonical product of $L^2(\mathbb{R}^4)$, which is invariant, provides a positive norm for this Hilbert space. Remarkably, the subsequent proper normalization of ``on-shell'' states directly ensures the usual $3d$  norm even in the presence of the external field, extending previous results derived through group averaging methods in the context of quantum gravity  \cite{ma.97,ma.00}.  

But in addition, the extended Hilbert space, normally considered as an auxiliary kinematic construction, is here promoted to a real physical space, in accordance with the consideration of time as an operator.  The importance of preserving the full $4d$ space becomes evident when the non-commutativity of the mass operators for different theories, e.g.  with and without external fields, is taken into account, which implies that the system naturally starts to explore the full space when an interaction is turned on. 
This approach also provides a $4d$ consistent Hilbert space for the Klein Gordon equation, which is explicitly covariant and hence differs from recent PaW treatments of square-root based Hamiltonian formulations \cite{KR.11} of the Klein-Gordon equation \cite{sm.19}. 
The present relativistic considerations have also allowed us to infer how to normalize states with infinite histories in a well defined non relativistic limit, providing a physical interpretation to the previous regularization proposal for the Schr\"odinger equation \cite{QT.15}. 
In this sense, PaW mechanism reveals to be particularly adequate for the relativistic context.

At the same time, the new features of the resulting formalism raise difficulties in the original relational interpretation \cite{PaW.83}: The time parameter ensuing from ``conditioning on the clock'' is unequivocally identified with time in a given frame of reference by the Klein Gordon equation. A relational interpretation would lead us to the conclusion that a single (quantum) clock is sufficient to describe the evolution of a particle for any observer, in clear contrast with the necessity of a synchronization convention such as Einstein synchronization \cite{E.05}. Moreover, this also requires the spectrum of $T$ to be continuous and unbounded so it can hardly be associated with an observable of a clock other than a coordinate. These considerations suggest that in this context it is more adequate to simply treat $t$ as an additional coordinate of the particle itself, as Stueckelberg approach also suggests (see Appendix \ref{B}). In the framework of general relativity, we would identify the time parameter with ``coordinate time'' rather than a time interval measured by a clock. 

On this basis, we have explored the construction of a Fock space where the building block is the particle as a 4-dimensional entity, extending thus the formalism to a many particle scenario and defining a ``second quantization'' of histories.
Through the definition of a proper subspace, an identification with the standard many particle  states follows. At the same time,  a richer time structure is revealed. This suggests a non-trivial extension to quantum field theory, different from a direct application of the original PaW mechanism. 
The present formalism may thus provide a novel perspective for dealing with different fundamental problems, like the concept of particles in curved space-times 
\cite{f.02}, the definition of a Hilbert space for the Wheeler DeWitt framework \cite{wdw.67,ku.11}, and the rigorous treatment of quantum correlations in time in quantum information and quantum optics \cite{p.04, k.19,s.13,z.19}.  
 
 \acknowledgments
We acknowledge support from CONICET (NLD, JMM) and  CIC (RR) of Argentina. Work supported by CONICET PIP Grant 112201501-00732.

\appendix
\section{Current density \label{A}}
In the present formalism, the Klein Gordon current density associated with an arbitrary state $|\Psi\rangle=\frac{1}{\sqrt{2\pi}}\int d^4x\,\psi(x)|x\rangle$ in the presence of a general potential $A_\mu(X)$,  
\begin{equation}j^\mu_A(x)=i\left(\psi^*(x)
D^\mu\psi(x)-\psi(x)D^{\mu*}\psi^*(x)\right)\,,
\label{jmu1}
\end{equation}
where $D^\mu=\partial^\mu+ie A^\mu$, can be written as 
\begin{equation}j^\mu(x)=2\pi\,\langle x|J^\mu_A|x\rangle \,,
\end{equation} where
\begin{equation}
J^\mu_A=-(P_A^\mu|\Psi\rangle\langle\Psi|+|\Psi\rangle\langle\Psi|P^\mu_A)\,,\label{jmupsi}\end{equation}
with $P^\mu_A=P^\mu+eA^\mu$,  is an hermitian operator. We can now express the 4-divergence of the current as 
\begin{eqnarray}\partial_\mu j^\mu_A(x)&=&2\pi i \,\langle x|[P_\mu,J^\mu_A]|x\rangle=
2\pi i\,\langle x|[P^A_\mu,J^\mu_A]|x\rangle\nonumber\\
&=&2\pi i\,\langle x|[|\Psi\rangle\langle\Psi|,{\cal J}_A]|x\rangle\label{j4}
\end{eqnarray}
where $P^A_\mu=P_\mu+eA_\mu$ and ${\cal J}_A=P^\mu_A P_\mu^A$ is the operator \eqref{JA}.  If $|\Psi\rangle$ is an eigenvector  of ${\cal J}_A$, i.e., a state with definite mass $|\Psi_{m^2}\rangle$, then $[|\Psi\rangle\langle\Psi|,{\cal J}_A]=0$ and we obtain the well known result 
\[\partial_\mu j^\mu_A(x)=0\,.\]
Previous relations can be immediately generalized to a two-state current density
\begin{eqnarray}j^\mu_A(\phi,\psi,x)&=&i\left(\phi^*(x)
D^\mu\psi(x)-\psi(x)D^{\mu *}\phi^*(x)\right)\nonumber\\
&=&2\pi\langle x|J^\mu_A(\Phi,\Psi)|x\rangle\,,
\label{jmu1a}
\end{eqnarray}
where
\begin{equation}
J^\mu_A(\Phi,\Psi)=-(P_A^\mu|\Psi\rangle\langle\Phi|+|\Psi\rangle\langle\Phi|P^\mu_A)\,.\label{jmupsia}\end{equation}
Eq.\ (\ref{j4}) now becomes 
\begin{eqnarray}\partial_\mu j^\mu_A(\phi,\psi,x)&=&
2\pi i\, \langle x|[|\Psi\rangle\langle\Phi|,{\cal J}_A]|x\rangle\label{j4a}\,.
\end{eqnarray} 
If $|\Psi\rangle$ and $|\Phi\rangle$ are both solutions of the Klein Gordon equation with the same mass, i.e., eigenstates of ${\cal J}_A$ with the same eigenvalue $m^2$, then $[|\Psi\rangle\langle\Phi|,{\cal J}_A]=0$, 
implying 
\begin{equation}\partial_\mu j^\mu_A(\phi,\psi,x)=0\,.\end{equation}
On the other hand,  for two eigenstates $|\Psi_{m^2}\rangle$, $|\Phi_{{m'}^2}\rangle$ with different masses   $m^2$ and ${m'}^2$,  we obtain instead 
\begin{equation}[|\Psi_{m^2}\rangle\langle\Phi_{{m'}^2}|,{\cal J}_A]=
({m'}^2-m^2)|\Psi_{m^2}\rangle\langle\Phi_{{m'}^2}|\,,
\end{equation}
implying the extended identity 
\begin{equation}\partial_\mu j^\mu_A(\phi_{m'^2},\psi_{{m}^2},x)=
i({m'}^2-m^2)\psi_{m^2}(x)\phi^*_{{m'}^2}(x)
\,,\label{djmu}\end{equation}
which holds for {\it any} mass-independent potential $A^\mu(X)$ (not necessarily time-independent). 

For $m^2={m'}^2$, integrating over $d^3x$ and assuming 
that $j^\mu_A(\phi_{m^2},\psi{_{m'}^2},x)$ vanishes for large $|{\bf x}|$, Eq.\ \eqref{djmu} leads to  
the well known result of $Q(\phi,\psi)$ constant in time, 
in agreement with Eq.\ \eqref{qpsi}. For general $m^2$, ${m'}^2$ this relation can be employed to re-obtain the previous relations \eqref{qpsi} and \eqref{mpm} (for a time and mass independent potential)  by integration of (\ref{djmu}) over $d^3x$, assuming again the vanishing of 
$j^\mu_A$ for large $|{\bf x}|$. 

The two-state current density can be also expressed as
\begin{equation}
    j^\mu_A(\phi,\psi,x)=\langle \Phi|J^\mu_A(x)|\Psi\rangle
    \end{equation}
where $J^\mu_A(x):=J^\mu_A(x,x)$ (Eq.\ \eqref{jmupsia})  is the hermitian operator 
\begin{equation}
   J^\mu_A(x)=-\left(\Pi(x) P_A^\mu+P_A^\mu\Pi(x)\right)\,,
    \end{equation}
with  $\Pi(x)=|x\rangle\langle x|$. 
Moreover, $Q_A(\phi,\psi)$ can be recast as 
\begin{eqnarray}
     Q_A(\phi,\psi)&=&\langle\Phi|\int d^3x\, J^\mu_A(x)|\Psi\rangle=\langle\Phi|Q_A(t)|\Psi\rangle\,,\;\;\;\;\;\;\\
     Q_A(t)&=&-\left(\Pi(t) P_A^0+P_A^0\Pi(t)\right)\,,
\end{eqnarray}
where $\Pi(t)=\int d^3x\,\Pi(x)=|t\rangle\langle t|\otimes\mathbbm{1}$. 
All relations of this appendix also hold of course in the free case $A_\mu(X)=0$ ($P^\mu_A\rightarrow P^\mu$, ${\cal J}_A\rightarrow {\cal J}$).  

\section{Extended quantization of a parameterized theory \label{B}}

Here we present an alternative version for deriving the history state formalism of a particle which is closely related to Stueckelberg proposal \cite{St.42}. While the previous approach is self-contained, this different perspective further clarifies that a relational interpretation is not strictly needed for the parameter $t$. At the same time, recovering the formalism in this way allows a first comparison with the conventional quantum gravity approach \cite{qg}. 

Consider the action of a one dimensional particle for a time independent Lagrangian
	\begin{equation}
	S\left[q(t)\right]=\int_{t_1}^{t_2}dt\, L(q,\dot{q})\,.
	\end{equation} 
	By promoting $t$ to a coordinate and parameterizing the configuration space $(t,q)$ with a variable $\tau$ we can write

	\begin{equation}
	S\left[q(\tau),t(\tau)\right]=\int_{\tau_1}^{\tau_2}d\tau\; \dot{t}L\left(q,\frac{\dot{q}}{\dot{t}}\right)\equiv \int_{\tau_1}^{\tau_2}d\tau\; \tilde{L}\left(q,\dot{q},\dot{t}\right)\,.
	\end{equation}
	
	The momenta associated with $\tilde{L}$ are \cite{qg}:
	\begin{eqnarray}\label{eq:supham}
	\tilde{p}_q&=&\frac{\partial \tilde{L}}{\partial \dot{q}}=p_q \nonumber\\
	{p}_t&=&\frac{\partial \tilde{L}}{\partial \dot{t}}=-H\,,
	\end{eqnarray}
	while the Hamiltonian is $\tilde{H}=\tilde{p}_q \dot{q}+{p}_t \dot{t}-\tilde{L}=\dot{t}(H+{p}_t)$. If we define the ``super Hamiltonian''
    $H_s\equiv H+\tilde{p}_t$ then from Eq.\ (\ref{eq:supham})
	\begin{equation}\label{eq:super}
	H_s=H+{p}_t\approx 0\,,
	\end{equation}
	where with $\approx$ we indicate this is a weak constraint \cite{d.58}. 
	By applying canonical quantization \emph{to the extended configuration space}, since $t$ and $p_t$ are in phase space, an enlarged Hilbert, which can be written as $\mathcal{H}=\mathcal{H}_t\otimes \mathcal{H}_q$, is obtained.
The super Hamiltonian constraint (\ref{eq:super}) defines the subspace 
\begin{equation}\label{eq:constr}
H_S|\Psi\rangle=(P_t\otimes \mathbb{I}+\mathbb{I}\otimes H)|\Psi\rangle=0\,,\end{equation}
which is precisely the universe equation of the PaW formalism for a one dimensional particle and continuum time discussed in Sec.\ \ref{sec:nonrel}. We have obtained by this method the familiar notions of the non relativistic history state formalism \emph{without considering a reference clock: $t$ is a coordinate.}

It should be stressed that the conventional quantization procedure of a parameterized system doesn't lead to the present formalism where a time operator is defined \cite{qg}. The key difference is that we are associating an enlarged Hilbert space to the extended configuration space such that the constraint (\ref{eq:constr}) has also a physical meaning instead of just a formal (or auxiliary) one \cite{ma.97}. The present proposal is more close to Stueckelberg approach to relativistic quantum mechanics \cite{St.42}. In fact, the Hamiltonian $R$ introduced by Stueckelberg under general relativistic considerations, which for a free particle reads $R=\frac{1}{2}P_\mu P^\mu$, leads to the Stueckelberg equation \cite{St.42}
\begin{equation}\label{eq:stuck}
    R \,\Psi(x,\tau)=i \frac{\partial}{\partial \tau}\Psi(x,\tau)\,,
\end{equation}
which for $\tau$ stationary solutions $\Psi(x,\tau)=\exp(\frac{-im^2}{2}\tau)\Psi(x)$ yields Eq.\ (\ref{eqgd}). The associated Stueckelberg norm $\int d^4x\, |\Psi(x,\tau)|^2$, which is $\tau$ independent for a  solution of Eq.\ (\ref{eq:stuck}), is precisely the one we have employed in Sec.\ \ref{sec:scalarp} and related to Klein Gordon norm for fixed mass solutions. The same considerations hold for the general case $R=\frac{1}{2}\pi_\mu \pi^\mu$, where $\pi_\mu=P_\mu+e A_\mu$, as follows immediately form the results of Sec.\ \ref{sec:ext}.


\begin{thebibliography}{48}%
\makeatletter
\providecommand \@ifxundefined [1]{%
 \@ifx{#1\undefined}
}%
\providecommand \@ifnum [1]{%
 \ifnum #1\expandafter \@firstoftwo
 \else \expandafter \@secondoftwo
 \fi
}%
\providecommand \@ifx [1]{%
 \ifx #1\expandafter \@firstoftwo
 \else \expandafter \@secondoftwo
 \fi
}%
\providecommand \natexlab [1]{#1}%
\providecommand \enquote  [1]{``#1''}%
\providecommand \bibnamefont  [1]{#1}%
\providecommand \bibfnamefont [1]{#1}%
\providecommand \citenamefont [1]{#1}%
\providecommand \href@noop [0]{\@secondoftwo}%
\providecommand \href [0]{\begingroup \@sanitize@url \@href}%
\providecommand \@href[1]{\@@startlink{#1}\@@href}%
\providecommand \@@href[1]{\endgroup#1\@@endlink}%
\providecommand \@sanitize@url [0]{\catcode `\\12\catcode `\$12\catcode
  `\&12\catcode `\#12\catcode `\^12\catcode `\_12\catcode `\%12\relax}%
\providecommand \@@startlink[1]{}%
\providecommand \@@endlink[0]{}%
\providecommand \url  [0]{\begingroup\@sanitize@url \@url }%
\providecommand \@url [1]{\endgroup\@href {#1}{\urlprefix }}%
\providecommand \urlprefix  [0]{URL }%
\providecommand \Eprint [0]{\href }%
\providecommand \doibase [0]{http://dx.doi.org/}%
\providecommand \selectlanguage [0]{\@gobble}%
\providecommand \bibinfo  [0]{\@secondoftwo}%
\providecommand \bibfield  [0]{\@secondoftwo}%
\providecommand \translation [1]{[#1]}%
\providecommand \BibitemOpen [0]{}%
\providecommand \bibitemStop [0]{}%
\providecommand \bibitemNoStop [0]{.\EOS\space}%
\providecommand \EOS [0]{\spacefactor3000\relax}%
\providecommand \BibitemShut  [1]{\csname bibitem#1\endcsname}%
\let\auto@bib@innerbib\@empty
\bibitem [{\citenamefont {Page}\ and\ \citenamefont {Wootters}(1983)}]{PaW.83}%
  \BibitemOpen
  \bibfield  {author} {\bibinfo {author} {\bibfnamefont {D.~N.}\ \bibnamefont
  {Page}}\ and\ \bibinfo {author} {\bibfnamefont {W.~K.}\ \bibnamefont
  {Wootters}},\ }\href@noop {} {\bibfield  {journal} {\bibinfo  {journal}
  {Phys. Rev. D}\ }\textbf {\bibinfo {volume} {27}},\ \bibinfo {pages} {2885}
  (\bibinfo {year} {1983})}\BibitemShut {NoStop}%
\bibitem [{\citenamefont {Gambini}\ \emph {et~al.}(2009)\citenamefont
  {Gambini}, \citenamefont {Porto}, \citenamefont {Pullin},\ and\ \citenamefont
  {Torterolo}}]{g.09}%
  \BibitemOpen
  \bibfield  {author} {\bibinfo {author} {\bibfnamefont {R.}~\bibnamefont
  {Gambini}}, \bibinfo {author} {\bibfnamefont {R.~A.}\ \bibnamefont {Porto}},
  \bibinfo {author} {\bibfnamefont {J.}~\bibnamefont {Pullin}}, \ and\ \bibinfo
  {author} {\bibfnamefont {S.}~\bibnamefont {Torterolo}},\ }\href@noop {}
  {\bibfield  {journal} {\bibinfo  {journal} {Phys. Rev. D}\ }\textbf {\bibinfo
  {volume} {79}},\ \bibinfo {pages} {041501(R)} (\bibinfo {year}
  {2009})}\BibitemShut {NoStop}%
\bibitem [{\citenamefont {Muga}\ and\ \citenamefont {Leavens}(2000)}]{m.00}%
  \BibitemOpen
  \bibfield  {author} {\bibinfo {author} {\bibfnamefont {J.~G.}\ \bibnamefont
  {Muga}}\ and\ \bibinfo {author} {\bibfnamefont {C.~R.}\ \bibnamefont
  {Leavens}},\ }\href@noop {} {\bibfield  {journal} {\bibinfo  {journal} {Phys.
  Rep.}\ }\textbf {\bibinfo {volume} {338}},\ \bibinfo {pages} {353} (\bibinfo
  {year} {2000})}\BibitemShut {NoStop}%
\bibitem [{\citenamefont {Giovannetti}\ \emph {et~al.}(2015)\citenamefont
  {Giovannetti}, \citenamefont {Lloyd},\ and\ \citenamefont {Maccone}}]{QT.15}%
  \BibitemOpen
  \bibfield  {author} {\bibinfo {author} {\bibfnamefont {V.}~\bibnamefont
  {Giovannetti}}, \bibinfo {author} {\bibfnamefont {S.}~\bibnamefont {Lloyd}},
  \ and\ \bibinfo {author} {\bibfnamefont {L.}~\bibnamefont {Maccone}},\
  }\href@noop {} {\bibfield  {journal} {\bibinfo  {journal} {Phys. Rev. D}\
  }\textbf {\bibinfo {volume} {92}},\ \bibinfo {pages} {045033} (\bibinfo
  {year} {2015})}\BibitemShut {NoStop}%
\bibitem [{\citenamefont {Boette}\ \emph {et~al.}(2016)\citenamefont {Boette},
  \citenamefont {Rossignoli}, \citenamefont {Gigena},\ and\ \citenamefont
  {Cerezo}}]{b.16}%
  \BibitemOpen
  \bibfield  {author} {\bibinfo {author} {\bibfnamefont {A.}~\bibnamefont
  {Boette}}, \bibinfo {author} {\bibfnamefont {R.}~\bibnamefont {Rossignoli}},
  \bibinfo {author} {\bibfnamefont {N.}~\bibnamefont {Gigena}}, \ and\ \bibinfo
  {author} {\bibfnamefont {M.}~\bibnamefont {Cerezo}},\ }\href@noop {}
  {\bibfield  {journal} {\bibinfo  {journal} {Phys. Rev. A}\ }\textbf {\bibinfo
  {volume} {93}},\ \bibinfo {pages} {062127} (\bibinfo {year}
  {2016})}\BibitemShut {NoStop}%
\bibitem [{\citenamefont {Moreva}\ \emph {et~al.}(2017)\citenamefont {Moreva},
  \citenamefont {Gramegna}, \citenamefont {Brida}, \citenamefont {Maccone},\
  and\ \citenamefont {Genovese}}]{m.17}%
  \BibitemOpen
  \bibfield  {author} {\bibinfo {author} {\bibfnamefont {E.}~\bibnamefont
  {Moreva}}, \bibinfo {author} {\bibfnamefont {M.}~\bibnamefont {Gramegna}},
  \bibinfo {author} {\bibfnamefont {G.}~\bibnamefont {Brida}}, \bibinfo
  {author} {\bibfnamefont {L.}~\bibnamefont {Maccone}}, \ and\ \bibinfo
  {author} {\bibfnamefont {M.}~\bibnamefont {Genovese}},\ }\href@noop {}
  {\bibfield  {journal} {\bibinfo  {journal} {Phys. Rev. D}\ }\textbf {\bibinfo
  {volume} {96}},\ \bibinfo {pages} {102005} (\bibinfo {year}
  {2017})}\BibitemShut {NoStop}%
\bibitem [{\citenamefont {Erker}\ \emph {et~al.}(2017)\citenamefont {Erker},
  \citenamefont {Mitchison}, \citenamefont {Silva}, \citenamefont {Woods},
  \citenamefont {Brunner},\ and\ \citenamefont {Huber}}]{e.17}%
  \BibitemOpen
  \bibfield  {author} {\bibinfo {author} {\bibfnamefont {P.}~\bibnamefont
  {Erker}}, \bibinfo {author} {\bibfnamefont {M.~T.}\ \bibnamefont
  {Mitchison}}, \bibinfo {author} {\bibfnamefont {R.}~\bibnamefont {Silva}},
  \bibinfo {author} {\bibfnamefont {M.~P.}\ \bibnamefont {Woods}}, \bibinfo
  {author} {\bibfnamefont {N.}~\bibnamefont {Brunner}}, \ and\ \bibinfo
  {author} {\bibfnamefont {M.}~\bibnamefont {Huber}},\ }\href@noop {}
  {\bibfield  {journal} {\bibinfo  {journal} {Phys. Rev. X}\ }\textbf {\bibinfo
  {volume} {7}},\ \bibinfo {pages} {031022} (\bibinfo {year}
  {2017})}\BibitemShut {NoStop}%
\bibitem [{\citenamefont {Dias}\ and\ \citenamefont {Parisio}(2017)}]{d.17}%
  \BibitemOpen
  \bibfield  {author} {\bibinfo {author} {\bibfnamefont {E.~O.}\ \bibnamefont
  {Dias}}\ and\ \bibinfo {author} {\bibfnamefont {F.}~\bibnamefont {Parisio}},\
  }\href@noop {} {\bibfield  {journal} {\bibinfo  {journal} {Phys. Rev. A}\
  }\textbf {\bibinfo {volume} {95}},\ \bibinfo {pages} {032133} (\bibinfo
  {year} {2017})}\BibitemShut {NoStop}%
\bibitem [{\citenamefont {Boette}\ and\ \citenamefont
  {Rossignoli}(2018)}]{b.18}%
  \BibitemOpen
  \bibfield  {author} {\bibinfo {author} {\bibfnamefont {A.}~\bibnamefont
  {Boette}}\ and\ \bibinfo {author} {\bibfnamefont {R.}~\bibnamefont
  {Rossignoli}},\ }\href@noop {} {\bibfield  {journal} {\bibinfo  {journal}
  {Phys. Rev. A}\ }\textbf {\bibinfo {volume} {98}},\ \bibinfo {pages} {032108}
  (\bibinfo {year} {2018})}\BibitemShut {NoStop}%
\bibitem [{\citenamefont {Mendes}\ and\ \citenamefont
  {Soares-Pinto}(2019)}]{m.18}%
  \BibitemOpen
  \bibfield  {author} {\bibinfo {author} {\bibfnamefont {L.~R.}\ \bibnamefont
  {Mendes}}\ and\ \bibinfo {author} {\bibfnamefont {D.~O.}\ \bibnamefont
  {Soares-Pinto}},\ }\href@noop {} {\bibfield  {journal} {\bibinfo  {journal}
  {Proc. Royal Soc. Lond A}\ }\textbf {\bibinfo {volume} {475}},\ \bibinfo
  {pages} {20190470} (\bibinfo {year} {2019})}\BibitemShut {NoStop}%
\bibitem [{\citenamefont {Martinelli}\ and\ \citenamefont
  {Soares-Pinto}(2019)}]{m.19}%
  \BibitemOpen
  \bibfield  {author} {\bibinfo {author} {\bibfnamefont {T.}~\bibnamefont
  {Martinelli}}\ and\ \bibinfo {author} {\bibfnamefont {D.~O.}\ \bibnamefont
  {Soares-Pinto}},\ }\href@noop {} {\bibfield  {journal} {\bibinfo  {journal}
  {Phys. Rev. A}\ }\textbf {\bibinfo {volume} {99}},\ \bibinfo {pages} {042124}
  (\bibinfo {year} {2019})}\BibitemShut {NoStop}%
\bibitem [{\citenamefont {Diaz}\ and\ \citenamefont
  {Rossignoli}(2019)}]{di.19}%
  \BibitemOpen
  \bibfield  {author} {\bibinfo {author} {\bibfnamefont {N.~L.}\ \bibnamefont
  {Diaz}}\ and\ \bibinfo {author} {\bibfnamefont {R.}~\bibnamefont
  {Rossignoli}},\ }\href@noop {} {\bibfield  {journal} {\bibinfo  {journal}
  {Phys. Rev. D}\ }\textbf {\bibinfo {volume} {99}},\ \bibinfo {pages} {045008}
  (\bibinfo {year} {2019})}\BibitemShut {NoStop}%
\bibitem [{\citenamefont {Smith}\ and\ \citenamefont {Ahmadi}(2019)}]{sm.19}%
  \BibitemOpen
  \bibfield  {author} {\bibinfo {author} {\bibfnamefont {A.~R.~H.}\
  \bibnamefont {Smith}}\ and\ \bibinfo {author} {\bibfnamefont
  {M.}~\bibnamefont {Ahmadi}},\ }\href@noop {} {\bibfield  {journal} {\bibinfo
  {journal} {Quantum}\ }\textbf {\bibinfo {volume} {3}},\ \bibinfo {pages}
  {160} (\bibinfo {year} {2019})};\ \bibinfo {note} {A.\ R.\ H.\  Smith and M.\ Ahmadi,
  arXiv:1904.12390 (2019).}\BibitemShut {Stop}%
\bibitem [{\citenamefont {Dirac}(1958)}]{d.58}%
  \BibitemOpen
  \bibfield  {author} {\bibinfo {author} {\bibfnamefont {P.~A.~M.}\
  \bibnamefont {Dirac}},\ }\href@noop {} {\bibfield  {journal} {\bibinfo
  {journal} {Proc. Royal Soc. Lond. A}\ }\textbf {\bibinfo {volume} {246}},\
  \bibinfo {pages} {333} (\bibinfo {year} {1958})}\BibitemShut {NoStop}%
\bibitem [{\citenamefont {Marolf}(1995{\natexlab{a}})}]{ma.95}%
  \BibitemOpen
  \bibfield  {author} {\bibinfo {author} {\bibfnamefont {D.}~\bibnamefont
  {Marolf}},\ }\href {\doibase 10.1088/0264-9381/12/10/007} {\bibfield
  {journal} {\bibinfo  {journal} {Class. Quantum Gravity}\ }\textbf {\bibinfo
  {volume} {12}},\ \bibinfo {pages} {2469} (\bibinfo {year}
  {1995}{\natexlab{a}})}\BibitemShut {NoStop}%
\bibitem [{\citenamefont {Marolf}(1995{\natexlab{b}})}]{ma2.95}%
  \BibitemOpen
  \bibfield  {author} {\bibinfo {author} {\bibfnamefont {D.}~\bibnamefont
  {Marolf}},\ }\href {\doibase 10.1088/0264-9381/12/5/011} {\bibfield
  {journal} {\bibinfo  {journal} {Class. Quantum Gravity}\ }\textbf {\bibinfo
  {volume} {12}},\ \bibinfo {pages} {1199–} (\bibinfo {year}
  {1995}{\natexlab{b}})}\BibitemShut {NoStop}%
\bibitem [{\citenamefont {Hartle}\ and\ \citenamefont {Marolf}(1997)}]{ma.97}%
  \BibitemOpen
  \bibfield  {author} {\bibinfo {author} {\bibfnamefont {J.~B.}\ \bibnamefont
  {Hartle}}\ and\ \bibinfo {author} {\bibfnamefont {D.}~\bibnamefont
  {Marolf}},\ }\href {\doibase 10.1103/PhysRevD.56.6247} {\bibfield  {journal}
  {\bibinfo  {journal} {Phys. Rev. D}\ }\textbf {\bibinfo {volume} {56}},\
  \bibinfo {pages} {6247} (\bibinfo {year} {1997})}\BibitemShut {NoStop}%
\bibitem [{\citenamefont {Marolf}(200)}]{ma.00}%
  \BibitemOpen
  \bibfield  {author} {\bibinfo {author} {\bibfnamefont {D.}~\bibnamefont
  {Marolf}},\ }in\ \href {\doibase 10.1142/9789812777386.0240} {\emph {\bibinfo
  {booktitle} {9th Marcel Grossmann Conference}}}\ (\bibinfo  {publisher}
  {World Scientific},\ \bibinfo {year} {200})\ pp.\ \bibinfo {pages}
  {1348--1349}\BibitemShut {NoStop}%
\bibitem [{\citenamefont {Kiefer}(2004)}]{qg}%
  \BibitemOpen
  \bibfield  {author} {\bibinfo {author} {\bibfnamefont {C.}~\bibnamefont
  {Kiefer}},\ }\href@noop {} {\bibfield  {journal} {\bibinfo  {journal} {Int.
  Ser. Monogr. Phys.}\ }\textbf {\bibinfo {volume} {136}},\ \bibinfo {pages}
  {1} (\bibinfo {year} {2007})}\BibitemShut {NoStop}%
\bibitem [{\citenamefont {Peres}(1999)}]{p.99}%
  \BibitemOpen
  \bibfield  {author} {\bibinfo {author} {\bibfnamefont {A.}~\bibnamefont
  {Peres}},\ }in\ \href@noop {} {\emph {\bibinfo {booktitle} {On Einstein's
  path}}}\ (\bibinfo  {publisher} {Springer},\ \bibinfo {year} {1999})\ pp.\
  \bibinfo {pages} {367--379}\BibitemShut {NoStop}%
\bibitem [{\citenamefont {Kucha{\v{r}}}(2011)}]{ku.11}%
  \BibitemOpen
  \bibfield  {author} {\bibinfo {author} {\bibfnamefont {K.~V.}\ \bibnamefont
  {Kucha{\v{r}}}},\ }\href@noop {} {\bibfield  {journal} {\bibinfo  {journal}
  {Int. J. Mod. Phys. D}\ }\textbf {\bibinfo {volume} {20}},\ \bibinfo {pages}
  {3} (\bibinfo {year} {2011})}\BibitemShut {NoStop}%
\bibitem [{\citenamefont {Bojowald}\ \emph
  {et~al.}(2011{\natexlab{a}})\citenamefont {Bojowald}, \citenamefont {Hoehn},\
  and\ \citenamefont {Tsobanjan}}]{b.11}%
  \BibitemOpen
  \bibfield  {author} {\bibinfo {author} {\bibfnamefont {M.}~\bibnamefont
  {Bojowald}}, \bibinfo {author} {\bibfnamefont {P.~A.}\ \bibnamefont {Hoehn}},
  \ and\ \bibinfo {author} {\bibfnamefont {A.}~\bibnamefont {Tsobanjan}},\
  }\href@noop {} {\bibfield  {journal} {\bibinfo  {journal} {Class. Quantum
  Gravity}\ }\textbf {\bibinfo {volume} {28}},\ \bibinfo {pages} {035006}
  (\bibinfo {year} {2011}{\natexlab{a}})}\BibitemShut {NoStop}%
\bibitem [{\citenamefont {Rovelli}(2011)}]{r.11}%
  \BibitemOpen
  \bibfield  {author} {\bibinfo {author} {\bibfnamefont {C.}~\bibnamefont
  {Rovelli}},\ }\href@noop {} {\bibfield  {journal} {\bibinfo  {journal}
  {Found. Phys.}\ }\textbf {\bibinfo {volume} {41}},\ \bibinfo {pages} {1475}
  (\bibinfo {year} {2011})}\BibitemShut {NoStop}%
\bibitem [{\citenamefont {Bojowald}\ \emph
  {et~al.}(2011{\natexlab{b}})\citenamefont {Bojowald}, \citenamefont
  {H{\"o}hn},\ and\ \citenamefont {Tsobanjan}}]{b.12}%
  \BibitemOpen
  \bibfield  {author} {\bibinfo {author} {\bibfnamefont {M.}~\bibnamefont
  {Bojowald}}, \bibinfo {author} {\bibfnamefont {P.~A.}\ \bibnamefont
  {H{\"o}hn}}, \ and\ \bibinfo {author} {\bibfnamefont {A.}~\bibnamefont
  {Tsobanjan}},\ }\href@noop {} {\bibfield  {journal} {\bibinfo  {journal}
  {Phys. Rev. D}\ }\textbf {\bibinfo {volume} {83}},\ \bibinfo {pages} {125023}
  (\bibinfo {year} {2011}{\natexlab{b}})}\BibitemShut {NoStop}%
\bibitem [{\citenamefont {Anderson}(2012)}]{a.12}%
  \BibitemOpen
  \bibfield  {author} {\bibinfo {author} {\bibfnamefont {E.}~\bibnamefont
  {Anderson}},\ }\href@noop {} {\bibfield  {journal} {\bibinfo  {journal} {Ann.
  Phys. (Berl.)}\ }\textbf {\bibinfo {volume} {524}},\ \bibinfo {pages} {757}
  (\bibinfo {year} {2012})}\BibitemShut {NoStop}%
\bibitem [{\citenamefont {Chataignier}()}]{chat.19}%
  \BibitemOpen
  \bibfield  {author} {\bibinfo {author} {\bibfnamefont {L.}~\bibnamefont
  {Chataignier}},\ }\href@noop {} {\ }\Eprint {http://arxiv.org/abs/1910.02998
  (2019)} {arXiv:1910.02998 (2019)} \BibitemShut {NoStop}%
\bibitem [{\citenamefont {Einstein}(1916)}]{e.16}%
  \BibitemOpen
  \bibfield  {author} {\bibinfo {author} {\bibfnamefont {A.}~\bibnamefont
  {Einstein}},\ }\href@noop {} {\bibfield  {journal} {\bibinfo  {journal} {Ann.
  Phys. (Berlin)}\ }\textbf {\bibinfo {volume} {354}},\ \bibinfo {pages} {769}
  (\bibinfo {year} {1916})}\BibitemShut {NoStop}%
\bibitem [{\citenamefont {Dirac}(1928)}]{di.28}%
  \BibitemOpen
  \bibfield  {author} {\bibinfo {author} {\bibfnamefont {P.~A.~M.}\
  \bibnamefont {Dirac}},\ }\href@noop {} {\bibfield  {journal} {\bibinfo
  {journal} {Proc. R. Soc. Lond. A}\ }\textbf {\bibinfo {volume} {117}},\
  \bibinfo {pages} {610} (\bibinfo {year} {1928})}\BibitemShut {NoStop}%
\bibitem [{\citenamefont {Klein}(1926)}]{K.26}%
  \BibitemOpen
  \bibfield  {author} {\bibinfo {author} {\bibfnamefont {O.}~\bibnamefont
  {Klein}},\ }\href {\doibase 10.1007/BF01397481} {\bibfield  {journal}
  {\bibinfo  {journal} {Z. Phys.}\ }\textbf {\bibinfo {volume} {37}},\ \bibinfo
  {pages} {895} (\bibinfo {year} {1926})}\BibitemShut {NoStop}%
\bibitem [{\citenamefont {Gordon}(1926)}]{G.26}%
  \BibitemOpen
  \bibfield  {author} {\bibinfo {author} {\bibfnamefont {W.}~\bibnamefont
  {Gordon}},\ }\href {\doibase 10.1007/BF01390840} {\bibfield  {journal}
  {\bibinfo  {journal} {Z. Phys.}\ }\textbf {\bibinfo {volume} {40}},\ \bibinfo
  {pages} {117} (\bibinfo {year} {1926})}\BibitemShut {NoStop}%
\bibitem [{\citenamefont {DeWitt}(1967)}]{wdw.67}%
  \BibitemOpen
  \bibfield  {author} {\bibinfo {author} {\bibfnamefont {B.~S.}\ \bibnamefont
  {DeWitt}},\ }\href@noop {} {\bibfield  {journal} {\bibinfo  {journal} {Phys.
  Rev.}\ }\textbf {\bibinfo {volume} {160}},\ \bibinfo {pages} {1113} (\bibinfo
  {year} {1967})}\BibitemShut {NoStop}%
\bibitem [{\citenamefont {Wigner}(1939)}]{w.39}%
  \BibitemOpen
  \bibfield  {author} {\bibinfo {author} {\bibfnamefont {E.}~\bibnamefont
  {Wigner}},\ }\href@noop {} {\bibfield  {journal} {\bibinfo  {journal} {Ann.
  of Math.}\ }\textbf {\bibinfo {volume} {40}},\ \bibinfo {pages} {149}
  (\bibinfo {year} {1939})}\BibitemShut {NoStop}%
\bibitem [{\citenamefont {Peskin}(2018)}]{p.18}%
  \BibitemOpen
  \bibfield  {author} {\bibinfo {author} {\bibfnamefont {M.~E.}\ \bibnamefont
  {Peskin}},\ }\href@noop {} {\emph {\bibinfo {title} {An introduction to
  quantum field theory}}}\ (\bibinfo  {publisher} {CRC Press, New York},\ \bibinfo {year}
  {2018})\BibitemShut {NoStop}%
\bibitem [{\citenamefont {Greiner}\ \emph {et~al.}(1990)\citenamefont {Greiner}
  \emph {et~al.}}]{Gr.90}%
  \BibitemOpen
  \bibfield  {author} {\bibinfo {author} {\bibfnamefont {W.}~\bibnamefont
  {Greiner}} \emph {et~al.},\ }\href@noop {} {\emph {\bibinfo {title}
  {Relativistic quantum mechanics}}},\ Vol.~\bibinfo {volume} {3}\ (\bibinfo
  {publisher} {Springer},\ \bibinfo {year} {1990})\BibitemShut {NoStop}%
\bibitem [{\citenamefont {Schwabl}(2008)}]{Sc.97}%
  \BibitemOpen
  \bibfield  {author} {\bibinfo {author} {\bibfnamefont {F.}~\bibnamefont
  {Schwabl}},\ }\href@noop {} {\emph {\bibinfo {title} {Advanced Quantum
  Mechanics}}}\ (\bibinfo  {publisher} {Springer},\ \bibinfo {year}
  {2008})\BibitemShut {NoStop}%
\bibitem [{\citenamefont {Maggiore}(2005)}]{M.05}%
  \BibitemOpen
  \bibfield  {author} {\bibinfo {author} {\bibfnamefont {M.}~\bibnamefont
  {Maggiore}},\ }\href@noop {} {\emph {\bibinfo {title} {A modern introduction
  to quantum field theory}}},\ Vol.~\bibinfo {volume} {12}\ (\bibinfo
  {publisher} {Oxford university press},\ \bibinfo {year} {2005})\BibitemShut
  {NoStop}%
\bibitem [{\citenamefont {Schwinger}(1951)}]{sc.51}%
  \BibitemOpen
  \bibfield  {author} {\bibinfo {author} {\bibfnamefont {J.}~\bibnamefont
  {Schwinger}},\ }\href@noop {} {\bibfield  {journal} {\bibinfo  {journal}
  {Phys. Rev.}\ }\textbf {\bibinfo {volume} {82}},\ \bibinfo {pages} {664}
  (\bibinfo {year} {1951})}\BibitemShut {NoStop}%
\bibitem [{\citenamefont {Bogolyubov}\ \emph {et~al.}(1975)\citenamefont
  {Bogolyubov}, \citenamefont {Logunov},\ and\ \citenamefont
  {Todorov}}]{Bog.75}%
  \BibitemOpen
  \bibfield  {author} {\bibinfo {author} {\bibfnamefont {N.~N.}\ \bibnamefont
  {Bogolyubov}}, \bibinfo {author} {\bibfnamefont {A.~A.}\ \bibnamefont
  {Logunov}}, \ and\ \bibinfo {author} {\bibfnamefont {I.~T.}\ \bibnamefont
  {Todorov}},\ }\href@noop {} {\emph {\bibinfo {title} {{Introduction to
  Axiomatic Quantum Field Theory}}}}\ (\bibinfo  {publisher} {W.A. Benjamin,
  Inc.},\ \bibinfo {year} {1975})\BibitemShut {NoStop}%
\bibitem [{\citenamefont {Unruh}\ and\ \citenamefont {Wald}(1989)}]{U.89}%
  \BibitemOpen
  \bibfield  {author} {\bibinfo {author} {\bibfnamefont {W.~G.}\ \bibnamefont
  {Unruh}}\ and\ \bibinfo {author} {\bibfnamefont {R.~M.}\ \bibnamefont
  {Wald}},\ }\href@noop {} {\bibfield  {journal} {\bibinfo  {journal} {Phys.
  Rev. D}\ }\textbf {\bibinfo {volume} {40}},\ \bibinfo {pages} {2598}
  (\bibinfo {year} {1989})}\BibitemShut {NoStop}%
\bibitem [{\citenamefont {Padmanabhan}\ and\ \citenamefont
  {Padmanabhan}(2011)}]{pa.11}%
  \BibitemOpen
  \bibfield  {author} {\bibinfo {author} {\bibfnamefont {H.}~\bibnamefont
  {Padmanabhan}}\ and\ \bibinfo {author} {\bibfnamefont {T.}~\bibnamefont
  {Padmanabhan}},\ }\href@noop {} {\bibfield  {journal} {\bibinfo  {journal}
  {Phys. Rev. D}\ }\textbf {\bibinfo {volume} {84}},\ \bibinfo {pages} {085018}
  (\bibinfo {year} {2011})}\BibitemShut {NoStop}%
\bibitem [{\citenamefont {Kowalski}\ and\ \citenamefont
  {Rembieli\ifmmode~\acute{n}\else \'{n}\fi{}ski}(2011)}]{KR.11}%
  \BibitemOpen
  \bibfield  {author} {\bibinfo {author} {\bibfnamefont {K.}~\bibnamefont
  {Kowalski}}\ and\ \bibinfo {author} {\bibfnamefont {J.}~\bibnamefont
  {Rembieli\ifmmode~\acute{n}\else \'{n}\fi{}ski}},\ }\href {\doibase
  10.1103/PhysRevA.84.012108} {\bibfield  {journal} {\bibinfo  {journal} {Phys.
  Rev. A}\ }\textbf {\bibinfo {volume} {84}},\ \bibinfo {pages} {012108}
  (\bibinfo {year} {2011})}\BibitemShut {NoStop}%
\bibitem [{\citenamefont {Einstein}(1905)}]{E.05}%
  \BibitemOpen
  \bibfield  {author} {\bibinfo {author} {\bibfnamefont {A.}~\bibnamefont
  {Einstein}},\ }\href@noop {} {\bibfield  {journal} {\bibinfo  {journal} {Ann.
  Phys. (Berlin)}\ }\textbf {\bibinfo {volume} {17}},\ \bibinfo {pages} {891}
  (\bibinfo {year} {1905})}\BibitemShut {NoStop}%
\bibitem [{\citenamefont {Ford}(2002)}]{f.02}%
  \BibitemOpen
  \bibfield  {author} {\bibinfo {author} {\bibfnamefont {L.}~\bibnamefont
  {Ford}},\ }in\ \href@noop {} {\emph {\bibinfo {booktitle} {General Relativity
  and Gravitation}}}\ (\bibinfo  {publisher} {World Scientific},\ \bibinfo
  {year} {2002})\ pp.\ \bibinfo {pages} {490--493}\BibitemShut {NoStop}%
\bibitem [{\citenamefont {Peres}\ and\ \citenamefont {Terno}(2004)}]{p.04}%
  \BibitemOpen
  \bibfield  {author} {\bibinfo {author} {\bibfnamefont {A.}~\bibnamefont
  {Peres}}\ and\ \bibinfo {author} {\bibfnamefont {D.~R.}\ \bibnamefont
  {Terno}},\ }\href@noop {} {\bibfield  {journal} {\bibinfo  {journal} {Rev.
  Mod. Phys}\ }\textbf {\bibinfo {volume} {76}},\ \bibinfo {pages} {93}
  (\bibinfo {year} {2004})}\BibitemShut {NoStop}%
\bibitem [{\citenamefont {Kull}\ \emph {et~al.}(2019)\citenamefont {Kull},
  \citenamefont {Gu{\'e}rin},\ and\ \citenamefont {Brukner}}]{k.19}%
  \BibitemOpen
  \bibfield  {author} {\bibinfo {author} {\bibfnamefont {I.}~\bibnamefont
  {Kull}}, \bibinfo {author} {\bibfnamefont {P.~A.}\ \bibnamefont
  {Gu{\'e}rin}}, \ and\ \bibinfo {author} {\bibfnamefont
  {{\v{C}}.}~\bibnamefont {Brukner}},\ }\href@noop {} {\bibfield  {journal}
  {\bibinfo  {journal} {npj Quantum Inf.}\ }\textbf {\bibinfo {volume} {5}},\
  \bibinfo {pages} {48} (\bibinfo {year} {2019})}\BibitemShut {NoStop}%
\bibitem [{\citenamefont {Shalm}\ \emph {et~al.}(2013)\citenamefont {Shalm},
  \citenamefont {Hamel}, \citenamefont {Yan}, \citenamefont {Simon},
  \citenamefont {Resch},\ and\ \citenamefont {Jennewein}}]{s.13}%
  \BibitemOpen
  \bibfield  {author} {\bibinfo {author} {\bibfnamefont {L.~K.}\ \bibnamefont
  {Shalm}}, \bibinfo {author} {\bibfnamefont {D.~R.}\ \bibnamefont {Hamel}},
  \bibinfo {author} {\bibfnamefont {Z.}~\bibnamefont {Yan}}, \bibinfo {author}
  {\bibfnamefont {C.}~\bibnamefont {Simon}}, \bibinfo {author} {\bibfnamefont
  {K.~J.}\ \bibnamefont {Resch}}, \ and\ \bibinfo {author} {\bibfnamefont
  {T.}~\bibnamefont {Jennewein}},\ }\href@noop {} {\bibfield  {journal}
  {\bibinfo  {journal} {Nat. Phys.}\ }\textbf {\bibinfo {volume} {9}},\
  \bibinfo {pages} {19} (\bibinfo {year} {2013})}\BibitemShut {NoStop}%
\bibitem [{\citenamefont {Zych}\ \emph {et~al.}(2019)\citenamefont {Zych},
  \citenamefont {Costa}, \citenamefont {Pikovski},\ and\ \citenamefont
  {Brukner}}]{z.19}%
  \BibitemOpen
  \bibfield  {author} {\bibinfo {author} {\bibfnamefont {M.}~\bibnamefont
  {Zych}}, \bibinfo {author} {\bibfnamefont {F.}~\bibnamefont {Costa}},
  \bibinfo {author} {\bibfnamefont {I.}~\bibnamefont {Pikovski}}, \ and\
  \bibinfo {author} {\bibfnamefont {{\v{C}}.}~\bibnamefont {Brukner}},\
  }\href@noop {} {\bibfield  {journal} {\bibinfo  {journal} {Nat. Commun.}\
  }\textbf {\bibinfo {volume} {10}},\ \bibinfo {pages} {3772} (\bibinfo {year}
  {2019})}\BibitemShut {NoStop}%
\bibitem [{\citenamefont {Stueckelberg}(1942)}]{St.42}%
  \BibitemOpen
  \bibfield  {author} {\bibinfo {author} {\bibfnamefont {E.~C.~G.}\
  \bibnamefont {Stueckelberg}},\ }\href@noop {} {\bibfield  {journal} {\bibinfo
   {journal} {Helv. Phys. Acta}\ }\textbf {\bibinfo {volume} {15}},\ \bibinfo
  {pages} {23} (\bibinfo {year} {1942})}\BibitemShut {NoStop}%
\end{thebibliography}

%
\end{document}